\documentclass[3p,authoryear,10pt]{elsarticle}

\usepackage{amssymb}
\usepackage{amsmath}
\usepackage{natbib}
\usepackage{subfigure}
\usepackage{graphicx}
\usepackage{epstopdf}
\usepackage{marvosym}
\usepackage{epsfig}
\usepackage{multirow}
\usepackage{tikz}
\usepackage{nicefrac}
\usepackage{arydshln}
\usepackage{url}
\usepackage{blkarray}
\usepackage{hyphenat} 
\usetikzlibrary{arrows,calc,intersections,positioning,arrows,decorations.pathmorphing,decorations.markings,shapes}%

\newcommand{\RN}[1]{%
  \textup{\uppercase\expandafter{\romannumeral#1}}%
}
\journal{Astrophysics and Space Science}

\begin{document}

\begin{frontmatter}

\title{
Propagation and Estimation of the Dynamical Behaviour of Gravitationally Interacting Rigid Bodies}

\author[TUDELFT]{D. Dirkx\corref{cor}}
\cortext[cor]{Corresponding author; Tel.: +31(0)15 2788866; Fax: +31(0)15 2781822 }
\ead{D.Dirkx@tudelft.nl}
\author[TUDELFT]{E. Mooij}
\author[TUDELFT]{B.C. Root}

\address[TUDELFT]{Delft University of Technology, Kluyverweg 1, 2629HS Delft, The Netherlands}


%

\begin{abstract}
Next-generation planetary tracking methods, such as interplanetary laser ranging (ILR) and same-beam interferometry (SBI) promise an orders-of-magnitude increase in the accuracy of measurements of solar system dynamics. This requires a reconsideration of modelling strategies for the translational and rotational dynamics of natural bodies, to ensure that model errors are well below the measurement uncertainties.

The influence of the gravitational interaction of the full mass distributions of celestial bodies, the so-called figure-figure effects, will need to be included for selected future missions. The mathematical {formulation of this problem to arbitrary degree} is often provided in an elegant and compact manner that is not trivially relatable to the formulation used in space geodesy and ephemeris generation. This complicates the robust implementation of such a model in operational software packages. We formulate the problem in a manner that is directly compatible with the implementation used in typical dynamical modelling codes:{ in terms of spherical harmonic coefficients and Legendre polynomials.} An analytical formulation for the associated variational equations for both translational and rotational motion is derived.

We apply our methodology to both Phobos and the KW4 binary asteroid system, to analyze the influence of figure-figure effects during estimation from next-generation tracking data. For the case of Phobos, omitting these effects during estimation results in{ relative errors of $0.42\%$ and $0.065\%$ for the $\bar{C}_{20}$ and $\bar{C}_{22}${ spherical harmonic gravity field coefficients}, respectively}. These values are below current uncertainties, but orders of magnitude larger than those obtained from past simulations for accurate tracking of a future Phobos lander, showing the need to apply the methodology outlined in this manuscript for selected future missions.

\end{abstract}


\end{frontmatter}

\section{Introduction}
\label{sec:introduction}

For the robust analysis of tracking data from planetary missions, the dynamics of solar system bodies under investigation should ideally be modelled to well below the {observational accuracy and precision. 
Several exceptionally accurate tracking-data} types are emerging for planetary missions, such as multi-wavelength radiometric range and Doppler measurements \citep{DehantEtAl2017}, same-beam interferometry (SBI) \citep{KikuchiEtAl2009,GregnaninEtAl2012}, and interplanetary laser ranging (ILR) \citep{Degnan2002,TuryshevEtAl2010,Dirkx2015c}. For the analysis of these data, dynamical models for natural bodies need to be developed and implemented to beyond the current state-of-the-art of typical state propagation and estimation software. 

{Examples of such software tools are GEODYN \citep{GenovaEtAl2016}, GINS \citep{MartyEtAl2009}, GMAT \citep{HughesEtAl2014}, NOE \citep{LaineyEtAl2004}, OREKIT \citep{Maisonobe2010} and Tudat (which we use in this manuscript, see \ref{app:tudat}). We stress that the full functionality of several of these codes (GMAT, OREKIT and Tudat being the exceptions) cannot be transparently determined, as up-to-date source code and documentation is not openly available for them. In this article we discuss, and present models to mitigate, one of the common challenges that these tools face for the analysis of future planetary tracking data.}

{The specific physical effects that must be incorporated for future missions depend strongly on the object under consideration, and the available tracking data types}. For both SBI and ILR, there is a need for sub-mm accurate dynamical models over the time span of the mission. For Doppler data, variations in range at the level of 1-10 $\mu$m/s need to be accounted for. {To meet the requirements that result from these tracking accuracies, various dynamical models may need to be improved, depending on the situation under consideration. Examples of such models include: a fully consistent tidal-rotational-translational dynamics model, realistic models for frequency-dependent tidal dissipation, detailed non-conservative force models for small bodies, and figure-figure gravitational interactions between massive bodies. The development of a model of the latter, for the purpose of state and parameter estimation, is the topic of the present paper.}

{Modelling barycentric motion to the mm-level} is presently limited by the knowledge of the properties of small solar system bodies, in particular main-belt asteroids \citep[\emph{e.g., }][]{KuchynkaEtAl2010}. However, the relative motion of solar system bodies in close proximity {(\emph{e.g.,} planetary satellite system or multiple asteroid system) is dominated by the gravity fields of these bodies themselves. Their relative motion is only weakly influenced by the gravitational fields of the other solar system bodies. As a result, the uncertainties in the local dynamics of such systems stem largely from uncertainties in, and mismodelling of the effects of, the gravitational interactions in the system itself. Measurements} of the local dynamics can be instrumental in improving the estimates of the properties of the bodies in the system  \citep[\textit{e.g.,} ][]{LaineyEtAl2007,FolknerEtAl2014,DirkxEtAl2016b}, provided that the dynamical model can be set up and parameterized to sufficient accuracy. 

Currently, the dynamical models of planetary/asteroid systems that are used in typical state propagation and estimation software cannot robustly capture the motion to the measurement accuracies of ILR and SBI \citep[\emph{e.g., }][]{Dirkx2015c, DehantEtAl2017}, which would prevent the data from being optimally exploited. Among others, characterizing such systems' dynamics will require a new level of detail for the models used to describe the gravitational interaction between extended bodies. Specifically, the coupling between higher-order terms in the gravity field expansions \citep[so-called figure-figure effects; ][]{BoisEtAl1992} will need to be included when propagating and estimating the translational and rotational dynamics of such bodies. {Although such models are incorporated in LLR data analysis software (although not necessarilly for arbitrary degree and order), the underlying models are not clearly described in literature, nor are these software frameworks openly available.}  {The resulting mathematical problem is also termed the} {full two-body problem (F2BP)}. The level to which these effects need to be included will depend strongly on the system under consideration. However, the \emph{a priori} assumption that such figure-figure terms can be neglected \citep[\emph{e.g.},][]{LaineyEtAl2007} will no longer be a given for many cases with high-accuracy, next-generation tracking systems. {At the very least, an evaluation of the magnitude of the influence of these terms should be made before performing the actual data processing}. 

The influence of low-order figure-figure terms on the translational and/or rotational dynamics of solar system bodies has been analyzed for a variety of cases, such as the Moon \citep{BoisEtAl1992,MullerEtAl2013}, Phobos \citep{BorderiesYoder1990,RambauxEtAl2012}, and binary asteroids \citep{FahnestockScheeres2008, HouEtAl2016}. Their analyses show that including figure-figure interactions is important for accurate dynamical modelling of selected systems of interacting bodies. For multiple asteroid systems, the higher-order gravitational interactions are especially strong, as a result of their highly irregular shapes and close orbits. As discussed by \cite{BatyginMordibelli2015}, understanding the spin interaction of these bodies is crucial in building a complete picture of the dynamical evolution of the solar system. {A body's rotational state is a key parameter in determining the influence of dissipative effects, which in turn play an important role in a body's long-term evolution.} 


A general formulation of mutual gravitational interaction potential of two extended bodies, which can be used to fully model such effects, was developed by \cite{Sidlichovsky1978} and \cite{Borderies1978}. Subsequently, \cite{Maciejewski1995} used these results to set up general translational and rotational equations of motion, later formally derived by \cite{LeeEtAl2007}, and extended to $N$ bodies by \cite{JiangEtAl2016}, including the static electric and magnetic potential. This method is described and applied further by \cite{MathisEtAl2009}, and \cite{CompereLemaitre2014}, using symmetric trace-free (STF) tensors \citep{Hartmann1994} and mass multipole moments. {Recently, an efficient representation of this problem was introduced by \cite{Boue2017} by applying angular momentum theory}. 
An equivalent formulation of the problem, in terms of inertia integrals instead of mass multipole moments, was developed by \cite{Paul1988}, \cite{Tricarico2008}, \cite{HouEtAl2016}, with a highly efficient implementation presented by \cite{Hou2018}. 
In an alternative approach, a formulation of the mutual interaction of homogeneous bodies is derived by \cite{WernerScheeres2005,FahnestockScheeres2006,HirabayashiScheeres2013} based on polyhedron shape models, which is highly valuable for the simulation of small bodies, such as binary asteroids \citep{FahnestockScheeres2008}. 

Explicit expansions of the mutual two-body interaction to low order have been derived by \cite{GiacagliaJefferys1971}, \cite{Schutz1981}, \cite{Ashenberg2007}, \cite{BoueLaskar2009}, and \cite{DobrovolskisKorycansky2013}, using a variety of approaches. For the analysis of future tracking-data types, the inclusion of {higher-order interactions effects} will be relevant, especially for highly non-spherical bodies in close orbits, such as binary asteroids \citep{HouEtAl2016, HouEtAl2017}.{ The need for figure-figure interactions in lunar rotational dynamics when analyzing LLR data is well known \citep{Eckhardt1981}. Recent re-analysis} by \cite{Hofmann2017} has shown the need to use the figure-figure interactions up to degree 3 in both the rotational and translational dynamics of the Earth-Moon system, for the analysis of modern LLR data. With the exception of LLR, figure-figure interactions have not been applied in tracking data analysis, and full algorithms to do so are not available in literature, nor is software to perform these analyses. Errors in dynamical modelling during data analysis can lead to biased estimates, and a true estimation error that is many times larger than the formal estimation error. 

The formulation of equations of motion in the F2BP does not trivially lend itself to the direct implementation in typical state propagation and estimation software tools. In such codes, the gravitational potential is described by the (normalized) spherical harmonic coefficients and Legendre polynomials, as opposed to the multipole moments/STF tensors and inertia integrals used in the F2BP. This gap between theoretical description and practical implementation must be closed before the figure-figure effects can be routinely included up to arbitrary degree in tracking-data analysis for (future) missions. Moreover, transparently and consistently including the figure-figure effects in a general manner in orbit determination and ephemeris generation algorithms has not yet been explored in detail. 

In this article, {the main goal is to derive} a direct link between the theoretical model for the F2BP and the implementation of the one-body potential, for both the propagation and estimation of the translational and rotational dynamics of the system. This will bridge the existing gap between theory and implementation in the context of spacecraft tracking and planetary geodesy. We start our development {in Section \ref{sec:potential}} from the formulation of \cite{MathisEtAl2009,CompereLemaitre2014} and \cite{Boue2017}, {and derive a direct and explicit link} with typical one-body implementations. {In Section \ref{sec:dynamicalEquations}, we present the equations of motion and derive the associated} variational equations, allowing the models to be used in orbit determination and ephemeris generation. A consistent formulation of the variational equations is crucial for the extraction of physical signatures from the (coupled) translational and rotational dynamics,  from tracking data. In Section \ref{sec:results}, we illustrate the impact of our method, by analyzing how estimation errors of the gravity field of Phobos, and the two bodies in the KW4 binary asteroid system, are affected if figure-figure interactions are omitted during the estimation.  Section \ref{sec:conclusions} summarizes the main results and findings. 

Our focus is on the development of an explicit link between the F2BP formulation and the formulations used in orbit determination/ephemeris generation, while ensuring a computational efficiency not prohibitive from a practical point of view. Our goal is not to improve the current state of the art in terms of computational performance \citep{Boue2017, Hou2018}.  

\section{Gravitational Potential}
\label{sec:potential}


We start by reviewing the formulation of the one-body potential in Section \ref{sec:oneBodyPotential}, followed by a discussion of the full two-body potential in Section \ref{sec:twoBodyPotential}. We discuss the transformation of the spherical harmonic coefficients between two reference frames in Section \ref{sec:coefficientTransformation}. Finally, we provide explicit expressions relating the computation of terms from the one-body potential to that of the full two-body potential in Section \ref{sec:implementationOneBody}. 

\subsection{Single-body Potential and Notation}
\label{sec:oneBodyPotential}
Applications in planetary geodesy typically represent the gravitational field of a single extended body by means of a spherical harmonic expansion of its gravitational potential \citep[\textit{e.g., }][]{MontenbruckGill2000,LaineyEtAl2004}:
\begin{align}
U(\mathbf{r})&=G\int_{B}\frac{dM}{|\mathbf{r}-\mathbf{s}|}\\
&=\frac{\mu}{r}\sum_{l=0}^{\infty}\sum_{m=0}^{l}\left(\frac{R}{r}\right)^{l}P_{lm}(\sin\varphi)\left(C_{lm}\cos m\vartheta+S_{lm}\sin m\vartheta\right)\label{eq:singleBodyPotentialReal}\\
&=\frac{\mu}{r}\sum_{l=0}^{\infty}\sum_{m=-l}^{l}\left(\frac{R}{r}\right)^{l}\mathcal{{M}}_{lm}Y_{lm}(\varphi,\vartheta)\label{eq:singleBodyPotentialComplex}\\
U_{lm}&=\frac{\mu}{r}\left(\frac{R}{r}\right)^{l}P_{lm}(\sin\varphi)\left(C_{lm}\cos m\vartheta+S_{lm}\sin m\vartheta\right)
\end{align}
where $\mathbf{r}$ denotes the position {at which the potential is evaluated and $\mathbf{s}$ denotes the position inside the body of the mass element $dM$. The spherical coordinates (radius, longitude, latitude) in a body-fixed frame are denoted by $r$, $\vartheta$ and $\varphi$. The reference radius of the body is denoted by $R$ and $\mu$ is the body's gravitational parameter. $P_{lm}$ and $Y_{lm}$ denote the unnormalized Legendre polynomials and spherical harmonic basis functions, respectively (both at degree $l$ and order $m$). {The term $U_{lm}$ is the full contribution from a single degree $l$ and order $m$ to the total potential.} ${\mathcal{M}_{lm}}$ represents the unnormalized mass multipole moments (typically used in the F2BP), and $C_{lm}$ and $S_{lm}$ are the spherical harmonic coefficients (typically used in spacecraft tracking analysis). The terms  $\mathcal{M}_{lm}$ are related to $C_{lm}$ and $S_{lm}$ as:
\begin{align}
\mathcal{M}_{lm}&=\begin{cases}
  \frac{(1+\delta_{0m})}{2}\left(C_{lm}-i S_{lm} \right), & m \ge 0 \\
  \mathcal{M}_{l,-m}^{*}(\text{-}1)^{m}\frac{(l-m)!}{(l+m)!}, & m < 0 \\
 \end{cases}\label{eq:multipoleMomentsPositiveNegative}
\end{align}
where $^{*}$ indicates the complex conjugate. The spherical harmonic basis functions in Eq. (\ref{eq:singleBodyPotentialComplex}) can be expressed as:
\begin{align}
Y_{lm}(\varphi,\vartheta)&= P_{lm}(\sin\varphi)e^{im\vartheta}\label{eq:unnormalizedBasisFunctions}\\
P_{lm}&=(\text{-}1)^{m}\frac{(l+m)!}{(l-m)!}P_{l,-m}\,\,(m<0)
\end{align}
Often the mass multipoles and basis functions are represented in a normalized manner. In this manuscipt, we will apply two normalizations: $4\pi$-normalized (for which quantities are represented with an overbar), and Schmidt semi-normalized (for which quantities are represented with a tilde). The Schmidt semi-normalized formulation is obtained from:
\begin{align}
\tilde{\mathcal{M}}_{lm}&=\frac{{\mathcal{M}}_{lm}}{\mathcal{\tilde{N}}_{lm}},\,\,\,\,\,\,\tilde{Y}_{lm}={{Y}_{lm}}{\mathcal{\tilde{N}}_{lm}}\label{eq:semiNormalization}\\
\mathcal{\tilde{N}}_{lm}&=(\text{-}1)^{m}\sqrt{\frac{2l+1}{4\pi}\frac{(l-m)!}{(l+m)!}}\label{eq:normalizationCoefficient}
\end{align}
which are used in the formulations of \cite{CompereLemaitre2014} and \cite{Boue2017}. In planetary geodesy, the  $4\pi$-normalized coefficients are typically used, for which:
\begin{align}
\bar{\mathcal{M}}_{lm}&=\frac{{\mathcal{M}}_{lm}}{\mathcal{\bar{N}}_{lm}},\,\,\,\,\,\,\bar{Y}_{lm}={{Y}_{lm}}{\mathcal{\bar{N}}_{lm}}\label{eq:fullNormalization}\\
\mathcal{\bar{N}}_{lm}&=\sqrt{\frac{(2-\delta_{0m})(2l+1)(l-m)!}{(l+m)!}}\label{eq:fullNormalizationCoefficient}
\end{align}
For the $4\pi$-normalized coefficients (which we shall simply refer to as 'normalized' from now on):
\begin{align}
\bar{\mathcal{M}}_{lm}&=\begin{cases}
\frac{(1+\delta_{0m})}{2}{(\bar{C}_{lm}-i\bar{S}_{lm})}, & m \ge 0 \\
(\text{-}1)^{m}\bar{\mathcal{M}}_{l,-m}^{*}, & m < 0 \\
\end{cases}\label{eq:normalizedMassMultipoles}\\
&=\mathcal{\bar{C}}_{lm}-i\mathcal{\bar{S}}_{lm}\label{eq:splitMonopole}\\
\bar{P}_{lm}&=\mathcal{\bar{N}}_{lm}P_{lm}\label{eq:geodesyNormalizedLegendre}\\
U&=\frac{\mu}{r}\sum_{l=0}^{\infty}\sum_{m=0}^{l}\left(\frac{R}{r}\right)^{l}\bar{P}_{lm}(\sin\varphi)\left(\bar{C}_{lm}\cos m\vartheta+\bar{S}_{lm}\sin m\vartheta\right)\label{eq:normalizedSingleBodyPotentialReal}
\end{align}
where we have introduced the $\mathcal{\bar{C}}_{lm}$, $\mathcal{\bar{S}}_{lm}$ notation (which we stress are distinct from $C_{lm}$ and $S_{lm}$) to avoid awkward expressions in later derivations.  {None of the final quantities that are needed in the computation are complex (Section \ref{sec:dynamicalEquations}).
However, the complex number notation is more concise, so we retain it in some sections to keep the derivation and analytical formulation tractable.}

\subsection{Two-body Interaction Potential}
\label{sec:twoBodyPotential}
For the interaction between extended bodies, we use the mutual force potential introduced by \cite{Borderies1978}. It is obtained from the following:
\begin{align}
V_{1\text{-}2}(\mathbf{r}_{1},\mathbf{r}_{2},\boldsymbol{\mathcal{R}}^{\nicefrac{\mathcal{F}_{2}}{I}}) =G\int_{B_{1}}\int_{B_{2}}\frac{dM_{1}dM_{2}}{d_{12}}\label{eq:mutualForcePotentialDoubleIntegral}
\end{align}
where $d_{12}$ denotes the distance between the mass elements $dM_{1}$ and $dM_{2}$ and $\mathbf{r}_{1}$ and $\mathbf{r}_{2}$ denote the inertial positions of the centers of mass of bodies 1 and 2, respectively. $B_{1}$ and $B_{2}$ denote full volume of bodies 1 and 2, respectively. The rotation from a frame $A$ to a frame $B$ is denoted as $\boldsymbol{\mathcal{R}}^{\nicefrac{B}{A}}$, while $\mathcal{F}_{i}$ denotes the frame fixed to body $i$ and $I$ denotes a given inertial frame (such as J2000).

The double integral in Eq. (\ref{eq:mutualForcePotentialDoubleIntegral}) can be expanded in terms of the mass multipoles and spherical harmonics \citep{Borderies1978}. We use a slightly modified\footnote{we use a $\sim$ to denote the semi-normalized parameters used by \cite{CompereLemaitre2014,Boue2017} to distinguish the terms from our (un)normalized formulations.} form of the notation used by \cite{MathisEtAl2009,CompereLemaitre2014}:
\begin{align}
V_{1\text{-}2}(\mathbf{r}^{\mathcal{F}_{1}},\boldsymbol{\mathcal{R}}^{\nicefrac{\mathcal{F}_{1}}{\mathcal{F}_{2}}})&=GM_{1}M_{2}\sum_{l_{1}=0}^{\infty}\sum_{m_{1}=-l_{1}}^{l_{1}}\sum_{l_{2}=0}^{\infty}\sum_{m_{2}=-l_{2}}^{l_{2}}(\text{-}1)^{l_{1}}\tilde{\gamma}_{l_{2},m_{2}}^{l_{1},m_{1}}R_{1}^{l_{1}}R_{2}^{l_{2}}\times\nonumber\\
&\hspace{2.5cm}\times\mathcal{\tilde{M}}_{l_{1},m_{1}}^{1,\mathcal{F}_{1}}\mathcal{\tilde{M}}_{l_{2},m_{2}}^{2,\mathcal{F}_{1}}(\boldsymbol{\mathcal{R}}^{\nicefrac{\mathcal{F}_{1}}{\mathcal{F}_{2}}})\frac{\tilde{Y}_{l_{1}+l_{2},m_{1}+m_{2}}(\vartheta,\varphi)}{r^{l_{1}+l_{2}+1}}\label{eq:complexMutualForcePotential}\\
&=GM_{1}M_{2}\sum_{l_{1}=0}^{\infty}\sum_{m_{1}=-l_{1}}^{l_{1}}\sum_{l_{2}=0}^{\infty}\sum_{m_{2}=-l_{2}}^{l_{2}} V_{l_{2},m_{2}}^{l_{1},m_{1}}(\mathbf{r}^{\mathcal{F}_{1}},\boldsymbol{\mathcal{R}}^{\nicefrac{\mathcal{F}_{1}}{\mathcal{F}_{2}}})\label{eq:complexMutualForcePotential2}
\end{align}
where {the distance between the centers of mass $\mathbf{r}_{21}=\mathbf{r}_{2}-\mathbf{r}_{1}$ is written as $\mathbf{r}$. The $\mathcal{F}_{1}$ superscript denotes that a vector is represented in the body-fixed frame of body 1. The angles $\vartheta$ and $\varphi$ denote the latitude and longitude of body 2, expressed in frame $\mathcal{F}_{1}$. 
The term $\tilde{\gamma}_{l_{2},m_{2}}^{l_{1},m_{1}}$ is a scaling term \citep{MathisEtAl2009}, {which can be written as:}
\begin{align}
\tilde{\gamma}_{l_{2},m_{2}}^{l_{1},m_{1}}&=\frac{\mathcal{\tilde{N}}_{l_{1},m_{1}}\mathcal{\tilde{N}}_{l_{2},m_{2}}}{\mathcal{\tilde{N}}_{l_{1}+l_{2},m_{1}+m_{2}}}\frac{(l_{1}+l_{2}-m_{1}-m_{2})!}{(l_{1}-m_{1})!(l_{2}-m_{2})!}
\end{align}
{from which it follows} that $\tilde{\gamma}_{l_{2},m_{2}}^{l_{1},m_{1}}=1$, if $l_{1}=0$ or $l_{2}=0$. 

As discussed in Section \ref{sec:introduction}, our goal is to find an explicit expression relating the implementation of Eq. (\ref{eq:complexMutualForcePotential}) to that of Eq. (\ref{eq:normalizedSingleBodyPotentialReal}), which uses $4\pi$-normalized mass multipoles. {Using  Eqs. (\ref{eq:unnormalizedBasisFunctions})-(\ref{eq:geodesyNormalizedLegendre})}, the terms $V_{l_{2},m_{2}}^{l_{1},m_{1}}$ can be rewritten explicitly as follows:
\begin{align}
V_{l_{2},m_{2}}^{l_{1},m_{1}}(\mathbf{r}^{\mathcal{F}_{1}},\boldsymbol{\mathcal{R}}^{\nicefrac{\mathcal{F}_{1}}{\mathcal{F}_{2}}})&=\bar{\gamma}^{l_{1},m_{1}}_{l_{2},m_{2}}\left(\frac{R_{1}}{r}\right)^{l_{1}}\left(\frac{R_{2}}{r}\right)^{l_{2}}\mathcal{\bar{M}}_{l_{1},m_{1},l_{2},m_{2}}^{1,2;\mathcal{F}_{1}}\bigg(\cos(|m_{1}+m_{2}|\vartheta)+...\nonumber\\
&\hspace{-1.0cm}...+i\left(s_{m_{1}+m_{2}}\sin(|m_{1}+m_{2}|\vartheta)\right)\bigg)\frac{\bar{P}_{l_{1}+l_{2},|m_{1}+m_{2}|}(\sin\varphi)}{r}\label{eq:realMutualForcePotentialNormalizedComponent}\\
\bar{\gamma}^{l_{1},m_{1}}_{l_{2},m_{2}}&=(\text{-}1)^{l_{1}}\tilde{\gamma}^{l_{1},m_{1}}_{l_{2},m_{2}}\sqrt{\frac{4\pi(2-\delta_{0m_{1}})(2-\delta_{0m_{2}})}{(2-\delta_{0(m_{1}+m_{2})})}}\sigma_{m_{1}+m_{2}}\label{eq:GammaBarTermDefinition}\\
&\sigma_{m}=\begin{cases}
   1, & m \ge 0 \\
   (\text{-}1)^{m}, & m < 0 \\
 \end{cases},\,\,\,\,\,\,\,\,\,s_{m}=\text{sgn}(m)\label{eq:SigmaTermDefinition}\\
&\mathcal{\bar{M}}_{l_{1},m_{1},l_{2},m_{2}}^{1,2;\mathcal{F}_{1}}=\mathcal{\bar{M}}_{l_{1},m_{1}}^{1,\mathcal{F}_{1}}\mathcal{\bar{M}}_{l_{2},m_{2}}^{2,\mathcal{F}_{1}}
\end{align}
Here, we have introduced effective two-body multipole moments $\mathcal{\bar{M}}_{l_{1},m_{1},l_{2},m_{2}}^{1,2;\mathcal{F}_{1}}$, defined by:
\begin{align}
\mathcal{\bar{M}}_{l_{1},m_{1},l_{2},m_{2}}^{1,2;\mathcal{F}_{1}}&=\mathcal{\bar{M}}_{l_{1},m_{1}}^{1,\mathcal{F}_{1}}\mathcal{\bar{M}}_{l_{2},m_2}^{2,\mathcal{F}_{1}}=\frac{\mathcal{\tilde{N}}_{l_{1}m_{1}}\mathcal{\tilde{N}}_{l_{2}m_{2}}}{\mathcal{\bar{N}}_{l_{1}m_{1}}\mathcal{\bar{N}}_{l_{2}m_{2}}}\mathcal{\tilde{M}}_{l_{1},m_{1}}^{1,\mathcal{F}_{1}}\mathcal{\tilde{M}}_{l_{2},m_2}^{2,\mathcal{F}_{1}}\\
&=\left(\mathcal{\bar{C}}_{l_{1},m_{1}}^{1,\mathcal{F}_{1}}\mathcal{\bar{C}}_{l_{2},m_{2}}^{2,\mathcal{F}_{1}}-\mathcal{\bar{S}}_{l_{1},m_{1}}^{1,\mathcal{F}_{1}} \mathcal{\bar{S}}_{l_{2},m_{2}}^{2,\mathcal{F}_{1}}\right)-i\left(\mathcal{\bar{C}}_{l_{1},m_{1}}^{1,\mathcal{F}_{1}} \mathcal{\bar{S}}_{l_{2},m_{2}}^{2,\mathcal{F}_{1}}+\mathcal{\bar{S}}_{l_{1},m_{1}}^{1,\mathcal{F}_{1}}\mathcal{\bar{C}}_{l_{2},m_{2}}^{2,\mathcal{F}_{1}}\right)\label{eq:twwBodyMultipoleCoefficientRelation}
\end{align}
The moments are expressed in the frame of body 1, and are therefore dependent on $\boldsymbol{\mathcal{R}}^{\nicefrac{\mathcal{F}_{1}}{\mathcal{F}_{2}}}$ if $l_{2}>0$. 

{The real part of the formulation for $V_{l_{2},m_{2}}^{l_{1},m_{1}}$ in Eq. (\ref{eq:realMutualForcePotentialNormalizedComponent}) is similar to a single  term $U_{lm}$ of the one-body potential in Eq. (\ref{eq:singleBodyPotentialReal}). 
Consequently, this formulation lends itself to the implementation in typical state propagation and estimation software (see Section \ref{sec:introduction}) by the correct change of variables, as we will discuss in detail in Section \ref{sec:implementationOneBody}. In later sections, the following decomposition for $V^{l_{1},m_{1}}_{l_{2},m_{2}}$ will ease some derivations:
\begin{align}
V^{l_{1},m_{1}}_{l_{2},m_{2}}&=\mathcal{\bar{M}}_{l_{2},m_{2}}^{2,\mathcal{F}_{1}}(\boldsymbol{\mathcal{R}}^{\nicefrac{\mathcal{F}_{1}}{\mathcal{F}_{2}}})\bar{u}^{l_{1},m_{1}}_{l_{2},m_{2}} \frac{\bar{Y}_{l_{1}+l_{2},m_{1}+m_{2}}(\vartheta,\varphi)}{r^{l_{1}+l_{2}+1}}\label{eq:singleMutualPotentialTermDecomposition}\\
\bar{u}^{l_{1},m_{1}}_{l_{2},m_{2}}&= \bar{\gamma}_{l_{2},m_{2}}^{l_{1},m_{1}}R_{1}^{l_{1}}R_{2}^{l_{2}}\mathcal{\bar{M}}_{l_{1},m_{1}}^{1,\mathcal{F}_{1}}
\end{align}
which explicitly separates the dependency on $\boldsymbol{\mathcal{R}}^{\nicefrac{\mathcal{F}_{1}}{\mathcal{F}_{2}}}$ and $\mathbf{r}^{\mathcal{F}_{1}}$.

{We assume that the mass multipoles $\mathcal{\bar{M}}^{i,\mathcal{F}_{i}}$ are time-inde\-pen\-dent (in their local frames $\mathcal{F}_{i}$). In principle, the inclusion of tidal effects \citep{MathisEtAl2009} does not fundamentally change the formulation of the mutual force potential. However, it does make} the $\mathcal{M}^{i,\mathcal{F}_{i}}_{lm}$ terms dependent on the relative positions and orientations of the bodies, substantially complicating the analytical formulation of the derivatives of these terms w.r.t. position and orientation {(Section \ref{sec:dynamicalEquations}). Therefore}, we limit ourselves to static gravity fields in this article, focussing on the relation between the one-body and two-body potential.


\subsection{Transformation of the Gravity Field Coefficients}
\label{sec:coefficientTransformation}
The main complication of using the mutual force potential {in Eq. (\ref{eq:complexMutualForcePotential}) lies in the orientation dependency of $\mathcal{\bar{M}}_{lm}^{2,\mathcal{F}_{1}}$}. Determining these values requires a transformation of multipole moments $\mathcal{\bar{M}}_{lm}^{2}$ from $\mathcal{F}_{2}$ (in which they are typically defined) to $\mathcal{F}_{1}$. A transformation from the semi-normalized multipoles $\mathcal{\tilde{M}}_{lm}^{2,\mathcal{F}_{2}}$ to $\mathcal{\tilde{M}}_{lm}^{2,\mathcal{F}_{1}}$ is given by \cite{Boue2017}, based on the methods from \cite{Wigner1959}, discussed in detail by \cite{VarshalovichEtAl1988}:
\begin{align}
\mathcal{\tilde{M}}_{lm}^{2,\mathcal{F}_{1}}=\sum_{k=-l}^ {l}D^{l}_{mk}(\boldsymbol{\mathcal{R}}^{\mathcal{F}_{1}/\mathcal{F}_{2}})\mathcal{\tilde{M}}_{lk}^{2,\mathcal{F}_{2}}\label{eq:normalizedCoefficientsTransformation}
\end{align}
where $D^{l}_{mk}$ represents the Wigner D-matrix of degree $l$ (with $-l\le m, k \le l$). Expressions for $D^{l}_{mk}$ can be found in literature in terms of Euler angles and Cayley-Klein parameters (among others). Here, we choose to express it in terms of the non-singular Cayley-Klein parameters, defined by two complex parameters ${a}$ and ${b}$, which are closely related to the unit quaternion more typically used in celestial mechanics (\ref{app:cayleyKlein}). We denote the vector containing the four elements of ${a}$ and ${b}$ as $\mathbf{c}$:
\begin{align}
\mathbf{c}=[\Re({a}),\, \Im({a}),\, \Re({b}),\, \Im({b})]^{T}\label{eq:ckVector}
\end{align}

We follow the same computational scheme as \cite{Boue2017} to determine the Wigner D matrices, which is a recursive formulation based on \cite{GimbutasGreengard2009}. Analytical formulations for $D^{0}_{mk}$ and  $D^{1}_{mk}$ are given in terms of ${a}$ and ${b}$ by \cite{VarshalovichEtAl1988} and \cite{Boue2017} as: 
\begin{align}
D^{0}_{0,0}=1,\,\,\,\,\,D^{1}_{m,k}=\begin{blockarray}{cccc}
&^{-1} & ^{0} & ^{1} \\
\begin{block}{c(ccc)}
  ^{-1} & (a^{*})^{2} & \sqrt{2}a^{*}b & b^{2} \\
  ^{0} & -\sqrt{2}a^{*}b^{*}  & |a|^{2}-|b|^{2} & \sqrt{2}ab \\
  ^{1} & (b^{*})^{2} & -\sqrt{2}ab^{*} & a^{2} \\
\end{block}
\end{blockarray}\label{eq:wignerRecursionInitialization}
\end{align} 
The following recursive formulation is then applied for $m\ge 0$ and $l\ge 2$:
\begin{align}
D^ {l}_{mk}&=\sum_{p=-1}^ {1}c_{mk}^{l;p}D^{1}_{1,-p}D^{l-1}_{m-1,k+p}\label{eq:wignerRecursive}\\
c_{mk}^{l;-1}&=\sqrt{\frac{(l+k)(l+k-1)}{(l+m)(l+m-1)}}\\
c_{mk}^{l;0}&=\sqrt{\frac{2(l+k)(l+k)}{(l+m)(l+m-1)}}\\
c_{mk}^{l;0}&=\sqrt{\frac{(l-k)(l-k-1)}{(l+m)(l+m-1)}}
\end{align} 
where $D^{l}_{mk}=0$ if $|m|>l$ or $|k|>l$. For $m<0$:
\begin{align}
D^ {l}_{mk}=(-1)^{m-k}\left(D^{l}_{-m,-l}\right)^{*}\label{eq:wignerDSymmetry}
\end{align} 

The transformation in Eq. (\ref{eq:normalizedCoefficientsTransformation}) {can be rewritten in terms of fully normalized multipole moments $\mathcal{\bar{M}}$ by using Eqs.  (\ref{eq:semiNormalization})-(\ref{eq:fullNormalizationCoefficient}) to obtain}: 
\begin{align}
{\mathcal{\bar{M}}}_{lm}^{2,\mathcal{F}_{1}}&=\sum_{k=-l}^{l}\bar{\nu}_{lmk}
D_{mk}^{l}{\mathcal{\bar{M}}}_{lk}^{2,\mathcal{F}_{2}}\label{eq:normalizedMultipoleTransformationFull}\\
\bar{\nu}_{lmk}&={(\text{-}1)^{k+m}}\sqrt{\frac{2-\delta_{0k}}{2-\delta_{0m}}}
\end{align}
Now, by virtue of Eq. (\ref{eq:normalizedMassMultipoles}}) {the transformation only needs to be performed for positive $m$, and we can rewrite Eq. (\ref{eq:normalizedMultipoleTransformationFull}) as:
\begin{align}
{\mathcal{\bar{M}}}_{lm}^{2,\mathcal{F}_{1}}&=\left(\bar{\nu}_{lm0}D^{l}_{m0}{\mathcal{\bar{M}}}_{l0}^{2,\mathcal{F}_{2}}+\sum_{k=1}^{k=l}\left(\bar{\nu}_{lmk}D^{l}_{mk}{\mathcal{\bar{M}}}_{lm}^{2,\mathcal{F}_{2}}+(-1)^{k}\bar{\nu}_{lm,-k}D^{l}_{m,-k}\left({\mathcal{\bar{M}}}_{lm}^{2,\mathcal{F}_{2}}\right)^{*}\right)\right)\label{eq:normalizedMultipoleTransformation}
\end{align}
{which allows for a direct} and transparent relation to spherical harmonic coefficients to be made (see Section \ref{sec:implementationOneBody}).

\subsection{Explicit Formulation in Terms of One-body Potential}
\label{sec:implementationOneBody}

Here, we explicitly provide equations linking the F2BP to the governing equations of the one-body potential in terms of spherical harmonics. 
From Eqs. (\ref{eq:normalizedMassMultipoles}) and  (\ref{eq:normalizedMultipoleTransformation}), the explicit equations for the transformed spherical harmonic coefficients become:
\begingroup
\allowdisplaybreaks[4]
\begin{align}
&\bar{\nu}_{lmk}D_{mk}^{l}=\Re_{mk}^{l}+i\Im_{mk}^{l}\label{eq:EfunctionDecomposition}\\
&\bar{C}^{2,\mathcal{F}_{1}}_{lm}=(2-\delta_{0m})\left(\Re^{l}_{m,0}\bar{C}_{l0}^{2,\mathcal{F}_{2}}+
\frac{1}{2}\sum_{k=1}^{l}\left(\left(\Re^{l}_{m,k}+(\text{-}1)^{k}\Re^{l}_{m,\text{-}k}
\right)\bar{C}^{2,\mathcal{F}_{2}}_{lk}+\left(\Im^{l}_{m,k}+(\text{-}1)^{k+1}\Im^{l}_{m,\text{-}k}
\right)\bar{S}^{2,\mathcal{F}_{2}}_{lk}\right)\right)\label{eq:transformedCCoefficientExplicit}\\
&\bar{S}^{2,\mathcal{F}_{1}}_{lm}=-(2-\delta_{0m})\left(\Im^{l}_{m,0}\bar{C}_{l0}^{2,\mathcal{F}_{2}}+
\frac{1}{2}\sum_{k=1}^{l}\left(\left(\Im^{l}_{m,k}+(\text{-}1)^{k}\Im^{l}_{m,\text{-}k}
\right)\bar{C}^{2,\mathcal{F}_{2}}_{lk}+\left( -\Re^{l}_{m,k}+(\text{-}1)^{k}\Re^{l}_{m,\text{-}k}
\right)\bar{S}^{2,\mathcal{F}_{2}}_{lk}\right)\right)\label{eq:transformedSCoefficientExplicit}
\end{align}
\endgroup
A single term $V_{l_{2},m_{2}}^{l_{1},m_{1}}$ of the mutual force potential coefficients can be computed using the exact same routines as those for computing a single term of the one body potential $U_{lm}$  by a correct substitution of variables. 
Comparing Eq. (\ref{eq:normalizedSingleBodyPotentialReal}) for $U_{lm}$ with the real part of  Eq. (\ref{eq:realMutualForcePotentialNormalizedComponent}) for $V_{l_{2},m_{2}}^{l_{1},m_{1}}$, we obtain, with Eq. (\ref{eq:twwBodyMultipoleCoefficientRelation}):
\begin{align}
l&\rightarrow l_{1}+l_{2}\label{eq:maximumEffectiveDegree}\\
m&\rightarrow |m_{1}+m_{2}|\\
\left(\frac{R}{r}\right)^{l}&\rightarrow\left(\frac{R_{1}}{r}\right)^{l_{1}}\left(\frac{R_{2}}{r}\right)^{l_{2}}\\
\bar{C}_{lm}&\rightarrow \bar{\gamma}^{l_{1},m_{1}}_{l_{2},m_{2}}\left(\mathcal{\bar{C}}_{l_{1},m_{1}}^{1,\mathcal{F}_{1}}\mathcal{\bar{C}}_{l_{2},m_{2}}^{2,\mathcal{F}_{1}}-\mathcal{\bar{S}}_{l_{1},m_{1}}^{1,\mathcal{F}_{1}} \mathcal{\bar{S}}_{l_{2},m_{2}}^{2,\mathcal{F}_{1}}\right)=\bar{C}_{l_{1,2};m_{1,2}}\label{eq:directCComponentTransformation}\\
\bar{S}_{lm}&\rightarrow s_{m_{1}+m_{2}}\bar{\gamma}^{l_{1},m_{1}}_{l_{2},m_{2}}\left(\mathcal{\bar{C}}_{l_{1},m_{1}}^{1,\mathcal{F}_{1}} \mathcal{\bar{S}}_{l_{2},m_{2}}^{2,\mathcal{F}_{1}}+\mathcal{\bar{S}}_{l_{1},m_{1}}^{1,\mathcal{F}_{1}}\mathcal{\bar{C}}_{l_{2},m_{2}}^{2,\mathcal{F}_{1}}\right)=\bar{S}_{l_{1,2};m_{1,2}}\label{eq:directSComponentTransformation}
\end{align}
with the $\mathcal{\bar{C}},\mathcal{\bar{S}}$ terms defined by Eq. (\ref{eq:splitMonopole}). In the above, we have introduced the $(\bar{C},\bar{S})_{l_{1,2};m_{1,2}}$ notation to denote the effective one-body spherical harmonic coefficients that are to be used to evaluate a single term $V_{l_{2},m_{2}}^{l_{1},m_{1}}$.
By substituting Eq. (\ref{eq:normalizedMassMultipoles}) into Eqs. (\ref{eq:directCComponentTransformation}) and (\ref{eq:directSComponentTransformation}) {we obtain (omitting the $\mathcal{F}_{1}$ superscripts)}:
\begingroup
\allowdisplaybreaks[3]
\begin{align}
 \bar{C}_{l_{1,2};m_{1,2}}&=\sigma_{m_{1}}\sigma_{m_{2}}\bar{\gamma}_{l_{2},m_{2}}^{{l_{1},m_{1}}}\frac{(1+\delta_{0m_{1}})(1+\delta_{0m_{2}})}{4}\left( \bar{C}_{l_{1},|m_{1}|}^{1} \bar{C}_{l_{2},|m_{2}|}^{2}-s_{m_{1}}s_{m_{2}} \bar{S}_{l_{1},|m_{1}|}^{1} \bar{S}_{l_{2},|m_{2}|}^{2} \right)
\label{eq:explicitEffectiveNormalizedCosineTerms}\\
  \bar{S}_{l_{1,2};m_{1,2}}&=s_{m_{1}+m_{2}}\sigma_{m_{1}}\sigma_{m_{2}}\bar{\gamma}_{l_{2},m_{2}}^{{l_{1},m_{1}}}
\frac{(1+\delta_{0m_{1}})(1+\delta_{0m_{2}})}{4}\left(s_{m_{2}} \bar{C}_{l_{1},|m_{1}|}^{1} \bar{S}_{l_{2},|m_{2}|}^{2}+s_{m_{1}} \bar{S}_{l_{1},|m_{1}|}^{1} \bar{C}_{l_{2},|m_{2}|}^{2} \right) 
\label{eq:explicitEffectiveNormalizedSineTerms}
\end{align} 
\endgroup
where the $s_{m}$ and $\sigma_{m}$ functions are defined in Eq. (\ref{eq:SigmaTermDefinition}). These equations provide the direct formulation of the effective two-body spherical harmonic coefficients in terms of the respective (transformed) one-body spherical harmonic coefficients: 
\begin{align}
V_{1\text{-}2}&=\frac{GM_{1}M_{2}}{r}\sum_{l_{1}=0}^{\infty}\sum_{m_{1}=-l_{1}}^{l_{1}}\sum_{l_{2}=0}^{\infty}\sum_{m_{2}=-l_{2}}^{l_{2}} \left(\frac{R_{1}}{r}\right)^{l_{1}}\left(\frac{R_{2}}{r}\right)^{l_{2}}P_{lm}(\sin\varphi)\left(\bar{C}_{l_{1,2};m_{1,2}}\cos m\vartheta+\bar{S}_{l_{1,2};m_{1,2}}\sin m\vartheta\right)\label{eq:mutualForcePotentialFromOneBody}
\end{align}
with $m={m_{1}+m_{2}}$ and $l=l_{1}+l_{2}$.

\subsection{Degree-two interactions - circular equatorial orbit}

To gain preliminary insight into the influence of figure-figure interactions, we perform a simplified analytical analysis of their effects. As a test case, we take two bodies in mutual circular equatorial orbits, with both bodies' rotations tidally locked to their orbit. In this configuration, the tidal bulges always lie along the same line, with a constant distance between the two bodies, and the frames $\mathcal{F}_{1}$ and $\mathcal{F}_{2}$ are equal, so that $(\bar{C},\bar{S})_{l,m}^{2,\mathcal{F}_{1}}=(\bar{C},\bar{S})_{l,m}^{2,\mathcal{F}_{2}}$. This represents a highly simplified model for, for instance, a binary asteroid system. 

Under these assumptions, we obtain the following equations from Section \ref{sec:implementationOneBody} (using $\bar{\gamma}^{l_{1},m_{1}}_{l_{2},m_{2}}=\bar{\gamma}^{l_{1},-m_{1}}_{l_{2},m_{2}}=\bar{\gamma}^{l_{1},m_{1}}_{l_{2},-m_{2}}$) for the relevant $V^{l_{1},m_{1}}_{l_{2},m_{2}}$ terms in Eq. (\ref{eq:complexMutualForcePotential2}) describing the interactions of $\bar{C}_{2,0}^{1}$ and $\bar{C}_{2,0}^{2}$, $\bar{C}_{2,0}^{1}$ and $\bar{C}_{2,2}^{2}$, and $\bar{C}_{2,2}^{1}$ and $\bar{C}_{2,2}^{2}$, respectively:
\begin{align}
V^{2,0}_{2,0}&=\bar{\gamma}^{2,0}_{2,0}\left(\frac{R_{1}}{r}\right)^{2}\left(\frac{R_{2}}{r}\right)^{2}\frac{\bar{C}_{2,0}^{1}\bar{C}_{2,0}^{2}P_{4,0}(\sin\varphi)}{r}\\
V^{2,0}_{2,2}+V^{2,0}_{2,-2}&=\bar{\gamma}^{2,0}_{2,2}\left(\frac{R_{1}}{r}\right)^{2}\left(\frac{R_{2}}{r}\right)^{2}\frac{\bar{C}_{2,0}^{1}\bar{C}_{2,2}^{2}P_{4,2}(\sin\varphi)\cos(2\vartheta)}{r}\label{eq:secondInteraction}\\
V^{2,2}_{2,2}+V^{2,2}_{2,-2}+V^{2,-2}_{2,2}+V^{2,-2}_{2,-2}+&=\frac{\bar{\gamma}^{2,2}_{2,2}}{2}\left(\frac{R_{1}}{r}\right)^{2}\left(\frac{R_{2}}{r}\right)^{2}\frac{\bar{C}_{2,2}^{1}\bar{C}_{2,2}^{2}\left(P_{4,0}(\sin\varphi)+P_{4,4}(\sin\varphi)\cos(4\vartheta)\right)}{r}
\end{align}
Note that the interaction between $\bar{C}_{2,2}^{1}$ and $\bar{C}_{2,0}^{2}$ can be obtained directly from Eq. (\ref{eq:secondInteraction}) by interchanging bodies 1 and 2.

Consequently, the figure-figure interactions of degree 2 present themselves in a similar manner as one-body interactions with $(l,m)$=(4,0), (4,2) and (4,4). As a result, the process of determining the relevance of figure-figure interactions is analogous, in our simplified situation, to determining relevance of one-body interactions of degree 4 terms, under the suitable change of variables for $\bar{C}_{4,0}$, $\bar{C}_{4,2}$ and $\bar{C}_{4,4}$. For more realistic cases (non-zero eccentricity and/or inclination), the effective spherical harmonic coefficients will be time-dependent. However, for low-eccentricity/inclination situations, the deviations will be relatively low, as long as bodies remain tidally locked, and consequently the frames $\mathcal{F}_{1}$ and $\mathcal{F}_{2}$ remain close.

The above analysis is also applicable in the case where the rotation of body 1 is \emph{not} tidally locked to body 2, so long as we set $\bar{C}_{22}^{1}$ to zero (no tidal bulge). Such a situation would be representative of a planetary satellite orbiting its host planet.

\section{Dynamical Equations}
\label{sec:dynamicalEquations}

{Here, we set up our equations} of translational and rotational motion in Section \ref{sec:eom}, following the approach of \cite{Maciejewski1995} and \cite{CompereLemaitre2014}. We use the algorithm by \cite{Boue2017} for the calculation of the torques, as it provides an efficient, elegant and non-singular implementation. Subsequently, we derive an analytical formulation for the variational equations in Section \ref{eq:variationalEquations}. We define the governing equations of the  dynamics of the bodies in an inertial frame, as opposed to the mutual position and orientation of the bodies that are used by \cite{Maciejewski1995} and \cite{CompereLemaitre2014}. Such an approach is more in line with typical implementation of few-body codes (\emph{e.g.}, solar system simulators; see Section \ref{sec:introduction}). {Also, it enables an easier implementation for the interaction of $N$ extended bodies.}

We provide the formulation {in which a quaternion defines} each body's orientation, instead of the full rotation matrices or Euler angles. {Quaternions are singularity-free, and their use was found by \cite{Fukushima2008} to be most efficient in terms of numerical error. 
We provide some basic aspects of quaternions in \ref{app:quatIntro}, and present the relations with Cayley-Klein parameters (see Section \ref{sec:coefficientTransformation}) in \ref{app:cayleyKlein}. 

\subsection{Equations of Motion}
\label{sec:eom}

To describe the complete two-body dynamics, we propagate the translational and rotational state of both bodies. Our state vector $\mathbf{x}$ is defined as follows:
\begin{align}
\mathbf{x}_{i}&=[\mathbf{r}_{i}\,\mathbf{v}_{i}\,\boldsymbol{\omega}_{i}^{\mathcal{F}_{i}}\,\mathbf{q}_{i}^{\nicefrac{I}{\mathcal{F}_{i}}}]^{T}\label{eq:singleBodyState}\\  
\mathbf{x}&=[\mathbf{x}_{1}\,\mathbf{x}_{2}]^{T}\label{eq:fullBodyState}
\end{align}
where $\mathbf{r}_{i}$ and $\mathbf{v}_{i}$ denote the inertial position and velocity of body $i$. The term $\boldsymbol{\omega}_{i}^{\mathcal{F}_{i}}$ denotes the angular velocity vector of body $i$ w.r.t. the $I$ frame, expressed in frame $\mathcal{F}_{i}$. The quaternion ${q}^{\nicefrac{I}{\mathcal{F}_{i}}}$ describes the quaternion rotation {operator from frame $\mathcal{F}_{i}$ to  frame $I$}. The variable $\mathbf{q}_{i}^{\nicefrac{I}{\mathcal{F}_{i}}}$ represents the vector containing the four entries of the quaternion (see \ref{app:quatIntro} for more details). 

In the remainder, we omit the superscripts for $\boldsymbol{\omega}_{i}^{\mathcal{F}_{i}}$ and $\mathbf{q}_{i}^{\nicefrac{I}{\mathcal{F}_{i}}}$ ({writing them as $\boldsymbol{\omega}_{i}$ and $\mathbf{q}_{i}$}), reintroducing an explicit frame notation only if it differs from the standard one in Eq. (\ref{eq:singleBodyState}). {We denote the quaternion that defines the full rotation as from $\mathcal{F}_{2}$ to $\mathcal{F}_{1}$ as $\mathbf{q}$.} 
We stress that neither the quaternion vector $\mathbf{q}$ nor the Cayley-Klein vector $\mathbf{c}$ (containing the entries of the complex numbers $a$ and $b$, see Eq. (\ref{eq:ckVector})) represents a vector in the typical use of the term (quantity with magnitude and direction). In this context we use the more general definition of {vector} used in computer science, \emph{i.e.}, a container of numerical values. 



Due to the symmetry of Eq. (\ref{eq:complexMutualForcePotential}) in $\mathbf{r}_{1}$ and $\mathbf{r}_{2}$, the translational equations of motion of the two bodies in an inertial frame are obtained immediately from \cite{Maciejewski1995} as:
\begin{align}
\dot{\mathbf{v}}_{1}
&=GM_{2}\boldsymbol{\mathcal{R}}^{\nicefrac{I}{\mathcal{F}_{1}}}\left(\sum_{l_{1}=0}^{\infty}\sum_{m_{1}=-l_{1}}^{l_{1}}\sum_{l_{2}=0}^{\infty}\sum_{m_{2}=-l_{2}}^{l_{2}} \frac{\partial}{\partial \mathbf{r}^{\mathcal{F}_{1}}}\left( V_{l_{2},m_{2}}^{l_{1},m_{1}}\left(\mathbf{r},\boldsymbol{\mathcal{R}}^{\nicefrac{\mathcal{F}_{1}}{\mathcal{F}_{2}}}\right)\right)\right)\label{eq:body1Translation}\\
\dot{\mathbf{v}}_{2}&=-\frac{M_{1}}{M_{2}}\dot{\mathbf{v}}_{1}\label{eq:body2Translation}
\end{align}
The mutual force potential depends on $\mathbf{r}_{i}$ through the spherical relative coordinates $r$, $\vartheta$ and $\varphi$, as shown in Eq. (\ref{eq:realMutualForcePotentialNormalizedComponent}), which is identical to the case for the single-body potential. Consequently, the calculation of the potential gradient can be done using standard techniques in space geodesy \citep{MontenbruckGill2000}, facilitated by our relation between the one-body and two-body potentials in Section \ref{sec:implementationOneBody}. 

{The rotational dynamics is described by \citep[\emph{e.g., }][]{Fukushima2008}}:
\begingroup
\allowdisplaybreaks[3]
\begin{align}
\dot{\boldsymbol{\omega}}_{i}^{\mathcal{F}_{i}}&=\mathbf{I}_{i}^{-1}\left(-\dot{\mathbf{I}}_{i}{\boldsymbol{\omega}}_{i}^{\mathcal{F}_{i}}+(\mathbf{I}_{i}{\boldsymbol{\omega}}_{i}^{\mathcal{F}_{i}})\times\boldsymbol{\omega}_{i}^{\mathcal{F}_{i}}+ \boldsymbol{M}_{i}^{\mathcal{F}_{i}}\right)\label{eq:bodyiRotationalDynamics}\\
\dot{\mathbf{q}}_{i}&=\mathbf{Q}(\mathbf{q}_{i})\boldsymbol{\omega_{i}}=\boldsymbol{\Omega}(\boldsymbol{\omega}_{i})\mathbf{q}_{i}\label{eq:body1RotationalKinematics}\\
\mathbf{Q}(\mathbf{q})&=\frac{1}{2}\begin{bmatrix} -q_{1} & -q_{2} & -q_{3}\\  q_{0} & -q_{3} & q_{2}\\ q_{3} & q_{0} & -q_{1}\\ -q_{2} & q_{1} & q_{0}\end{bmatrix}\hspace{0.5cm}\boldsymbol{\Omega}(\boldsymbol{\omega})=\frac{1}{2}\begin{bmatrix}0 & -\omega_{1}&-\omega_{2}&-\omega_{3}\\ \omega_{1} & 0&\omega_{3}&-\omega_{2}\\ \omega_{2}& -\omega_{3}&0&\omega_{1}\\ \omega_{3} & \omega_{2}&-\omega_{1}&0\end{bmatrix}
\end{align}
\endgroup
where {we take the inertia tensor $\mathbf{I}_{i}$ of body $i$} in coordinates fixed to body $i$.  For our application, we set $\dot{\mathbf{I}}=0$ (see Section \ref{sec:twoBodyPotential}). {The $\dot{\mathbf{q}}=\mathbf{Q}\boldsymbol{\omega}$ formulation is used for the numerical propagation.}


{For the computation of gravitational torques in the F2BP, \cite{Boue2017} has very recently introduced a novel approach to compute the terms $\mathbf{M}_{i}^{I}$, based on \cite{VarshalovichEtAl1988}, which is computationally more efficient, and allows the torque to be evaluated without resorting to Euler angles. We express a single term $V^{l_{1},m_{1}}_{l_{2},m_{2}}$ as in Eq. (\ref{eq:singleMutualPotentialTermDecomposition}), which allows for expressing $\mathbf{M}_{2}^{\mathcal{F}_{1}}$ as follows using the angular momentum operator $\boldsymbol{\hat{\mathcal{J}}}$:
\begin{align}
\mathbf{M}_{2}^{\mathcal{F}_{1}}&=-\boldsymbol{\hat{\mathcal{J}}}\left(V_{1\text{-}2}\right)\\
&=-GM_{1}M_{2}\sum_{l_{1}=0}^{\infty}\sum_{m_{1}=-l_{1}}^{l_{1}}\sum_{l_{2}=0}^{\infty}\sum_{m_{2}=-l_{2}}^{l_{2}}\boldsymbol{\hat{\mathcal{J}}}\left(\mathcal{\bar{M}}_{l_{2},m_{2}}^{2,\mathcal{F}_{1}}\right)\bar{u}^{l_{1},m_{1}}_{l_{2},m_{2}} \frac{Y_{l_{1}+l_{2},m_{1}+m_{2}}(\vartheta,\varphi)}{r^{l_{1}+l_{2}+1}}\label{eq:momentWithAngMomOp}\\
\boldsymbol{\hat{\mathcal{J}}}\left(\mathcal{\bar{M}}_{l_{2},m_{2}}^{2,\mathcal{F}_{1}}\right)&=\sum_{k_{2}=-l_{2}}^{l_{2}}\bar{\nu}_{lmk}\boldsymbol{\hat{\mathcal{J}}}\left(D_{m_{2},k_{2}}^{l_{2}}\right)\mathcal{\bar{M}}_{l_{2},k_{2}}^{2,\mathcal{F}_{2}}\label{eq:singleMultipoleAngularMomentumOperator}
\end{align}
The terms $\boldsymbol{\hat{\mathcal{J}}}\left(D_{m_{2},k_{2}}^{l_{2}}\right)$ can be evaluated directly in Cartesian coordinates, from the expressions given by \cite{Boue2017}:
\begin{align}
\boldsymbol{\hat{\mathcal{J}}}\left(D_{m,k}^{l}\right)&=\mathbf{K}_{lm}\hat{\mathbf{D}}^{l}_{m,k}\label{eq:angMomOnWigner}\\
\hat{\mathbf{D}}^{l}_{m,k}&=\begin{pmatrix}D^ {l}_{m+1,k}\\ D^ {l}_{m,k} \\ D^ {l}_{m-1,k}\end{pmatrix}\\
\mathbf{K}_{lm}&=\begin{pmatrix} \frac{\sqrt{l(l+1)-m(m+1)}}{2}i & 0 & -\frac{\sqrt{l(l+1)-m(m-1)}}{2}i \\
-\frac{\sqrt{l(l+1)-m(m+1)}}{2} & 0 & -\frac{\sqrt{l(l+1)-m(m-1)}}{2} \\
0 & -m & 0\\
 \end{pmatrix}\label{eq:angularMomentumOperatorTransform}
\end{align} 
where the terms $D^{l}_{m,k}$ are computed as described in Section \ref{sec:coefficientTransformation}.

The formulation of Eq. (\ref{eq:momentWithAngMomOp}) can be evaluated using the approach of Section \ref{sec:implementationOneBody}. The difference that needs to be introduced when evaluating $\boldsymbol{\hat{\mathcal{J}}}(V^{l_{1},m_{1}}_{l_{2},m_{2}})$, instead of $V^{l_{1},m_{1}}_{l_{2},m_{2}}$, is that Eq. (\ref{eq:EfunctionDecomposition}) is replaced by:
\begin{align}
&\bar{\nu}_{lmk}\boldsymbol{\hat{\mathcal{J}}}(D_{mk}^{l})=\boldsymbol{\Re}_{mk}^{l}+i\boldsymbol{\Im}_{mk}^{l}
\end{align}
subsequently replacing $({\Re},{\Im})_{mk}^{l}$ in Eq. (\ref{eq:transformedCCoefficientExplicit}) and (\ref{eq:transformedSCoefficientExplicit}) with $(\boldsymbol{\Re},\boldsymbol{\Im})_{mk}^{l}$ provides $\boldsymbol{\hat{\mathcal{J}}}(\bar{C}_{lm}^{2,\mathcal{F}_{1}})$ and $\boldsymbol{\hat{\mathcal{J}}}(\bar{S}_{lm}^{2,\mathcal{F}_{1}})$, respectively. Continuing in the same manner, Eqs. (\ref{eq:explicitEffectiveNormalizedCosineTerms}) and (\ref{eq:explicitEffectiveNormalizedSineTerms}) are adapted to obtain $\boldsymbol{\hat{\mathcal{J}}}(\bar{C}_{l_{1,2};m_{1,2}})$ and $\boldsymbol{\hat{\mathcal{J}}}(\bar{S}_{l_{1,2};m_{1,2}})$, respectively. Finally, we evaluate, analogously to Eq. (\ref{eq:mutualForcePotentialFromOneBody}):
\begin{align}
\boldsymbol{\hat{\mathcal{J}}}(V_{1\text{-}2})=\frac{-GM_{1}M_{2}}{rs}\sum_{l_{1}=0}^{\infty}\sum_{m_{1}=-l_{1}}^{l_{1}}&\sum_{l_{2}=0}^{\infty}\sum_{m_{2}=-l_{2}}^{l_{2}}\left(\frac{R_{1}}{r}\right)^{l_{1}}\left(\frac{R_{2}}{r}\right)^{l_{2}}P_{lm}(\sin\varphi)\cdot...\nonumber\\
&...\cdot\left(\boldsymbol{\hat{\mathcal{J}}}(\bar{C}_{l_{1,2};m_{1,2}})\cos m\vartheta+\boldsymbol{\hat{\mathcal{J}}}(\bar{S}_{l_{1,2};m_{1,2}})\sin m\vartheta\right)\label{eq:mutualForcePotentialAngMomOpFromOneBody}
\end{align}

To complete the rotational equations of motion, the torques $\mathbf{M}_{1}$ and $\mathbf{M}_{2}$ are related as:
\begin{align}
&\mathbf{M}_{1}^{\mathcal{F}_{1}}+\mathbf{M}_{2}^{\mathcal{F}_{1}}=\mathbf{r}^{\mathcal{F}{1}}\times\left(\frac{\partial V_{1\text{-}2}}{\partial \mathbf{r}^{\mathcal{F}{1}}}\right)^{T} \label{eq:moment1Formulation}\\
&\mathbf{M}_{2}^{\mathcal{F}_{2}}=\boldsymbol{\mathcal{R}}^{\nicefrac{\mathcal{F}_{2}}{\mathcal{F}_{1}}}\mathbf{M}_{2}^{\mathcal{F}_{1}} \label{eq:moment2FormulationInFrame1}
\end{align}
as a consequence of conservation of angular momentum,

\subsection{Variational Equations}
\label{eq:variationalEquations}

{To estimate the rotational and translational behavior in the F2BP from planetary tracking data, we need a formulation of the variational equations \citep{MontenbruckGill2000,SchutzEtAl2004,MilaniGronchi2010} for the dynamical model defined in Section \ref{sec:eom}. This requires the computation of the partial derivatives of the accelerations and torques w.r.t. the current state, as well as a parameter vector.  These partial derivatives can be computed numerically,} but an analytical formulation is often computationally more efficient and less prone to numerical error. This is especially true in the case of gravitational accelerations, for which the computation can be performed in a recursive manner based on the acceleration components. 

Typically, only the translational motion is estimated dynamically (\emph{i.e}, represented in the state vector $\mathbf{x}$) from the tracking data. The rotational behavior of the bodies is most often parameterized as a mean rotational axis and rate (possibly with slow time variations) in addition to a spectrum of libration amplitudes. {These spectra may be obtained from fitting observations of solar system bodies \citep{ArchinalEtAl2018}, or a numerical integration of the rotational equations of motion, as done by \cite{RambauxEtAl2012} for Phobos, based on current models for the physical properties of the system.} This approach partly decouples the translational and rotational dynamics in the estimation. A coupled determination of the initial rotational and translational state has been performed for only a limited number of cases, {most notably in the case of the Moon using LLR data \citep[\textit{e.g., }][]{NewhallWilliams1997,FolknerEtAl2014,Hofmann2017}}. Dynamical estimation of rotational motion of asteroid Bennu was performed by \cite{MazaricoEtAl2017} during a simulation study in preparation for the OSIRIS-REx mission. There, the possibly significant wobble of the asteroid is shown to require a dynamical approach to rotation characterization when performing accurate proximity operations.  We propose a similar approach here, in which {the full state vector $\mathbf{x}$ is} dynamically determined, and all couplings are automatically included in the estimation model.

We stress that the approach of estimating libration amplitudes, rotation rates, \emph{etc.}, is the preferred choice for analyses of data from most current solar system missions. The signatures of the full rotational behavior are typically too weak in these data sets to warrant the full dynamical estimation (with the clear exception of lunar rotation from LLR data). However, such a full dynamical approach was shown to be crucial by \cite{DirkxEtAl2014} for the realistic data analysis of a Phobos Laser Ranging mission, both to ensure full consistency between all estimated parameters, and to prevent the estimation of  an excessive number of correlated libration parameters. 

To determine the full state behavior as a function of time, the state $\mathbf{x}$ at some time $t_{0}$ (denoted $\mathbf{x}_{0}$), as well as a set of parameters represented here as $\mathbf{p}$, are estimated. The influence of initial state and parameter errors are mapped to any later time by means of the state transition and sensitivity matrices, $\boldsymbol{\Phi}(t,t_{0})$ and $S(t)$, defined as:
\begin{align}
\boldsymbol{\Phi}(t,t_{0})&=\frac{\partial\mathbf{x}(t)}{\partial \mathbf{x}_{0}}\\
\mathbf{S}(t)&=\frac{\partial{\mathbf{x}}}{\partial\mathbf{p}}
\end{align}
The differential equations governing the time-behavior of these matrices are given by:
\begin{align}
\dot{\boldsymbol{\Phi}}(t,t_{0})&=\frac{\partial \dot{\mathbf{x}}}{\partial\mathbf{x}}\boldsymbol{\Phi}(t,t_{0})\label{eq:varEq}\\
\dot{\mathbf{S}}(t)&=\frac{\partial\dot{\mathbf{x}}}{\partial\mathbf{x}}\mathbf{S}(t)+\frac{\partial\dot{\mathbf{x}}}{\partial\mathbf{p}}\label{eq:varEqSens}
\end{align}
where the state derivative model $\mathbf{\dot{x}}$ is defined by Eqs. (\ref{eq:body1Translation})-(\ref{eq:body1RotationalKinematics}). In Sections \ref{sec:stateTransition} and \ref{sec:sensitivity} we present the detailed formulation of the terms in Eqs (\ref{eq:varEq}) and (\ref{eq:varEqSens}), respectively.

\subsubsection{State transition matrix}
\label{sec:stateTransition}
Writing out the partial derivatives in Eq. (\ref{eq:varEq}), we obtain four matrix blocks of the following structure (with $i,j=1...2)$:

\begingroup
\renewcommand*{\arraystretch}{1.5}
\begin{align}
\frac{\partial{\dot{\mathbf{x}}}_{i}}{\partial\mathbf{x}_{j}}=\begin{pmatrix}\mathbf{0}_{3\times 3} & \delta_{ij}\mathbf{1}_{3\times 3} & \mathbf{0}_{3\times 4}  & \mathbf{0}_{3\times 3} \\
\frac{\partial\dot{\mathbf{v}}_{i}}{\partial\mathbf{r}_{j}} & \mathbf{0}_{3\times 3}  & \frac{\partial\dot{\mathbf{v}}_{i}}{\partial\mathbf{q}_{j}} & \mathbf{0}_{3\times 3}\\
\mathbf{0}_{4\times 3} & \mathbf{0}_{4\times 3} & \delta_{ij}\boldsymbol{\Omega}(\boldsymbol{\omega}_{i}) & \delta_{ij}\mathbf{Q}(\mathbf{q}_{i})\\
\frac{\partial\dot{\boldsymbol{\omega}}_{i}}{\partial\mathbf{r}_{j}} & \mathbf{0}_{3\times 3}  & \frac{\partial\dot{\boldsymbol{\omega}}_{i}}{\partial\mathbf{q}_{j}} & \delta_{ij}\frac{\partial\dot{\boldsymbol{\omega}}_{i}}{\partial\boldsymbol{\omega}_{j}}
\end{pmatrix}\label{eq:stateDerivativePartialMatrix}
\end{align}
\endgroup
In this section, we will explicitly derive equations {that enable an analytical evaluation of the partial derivatives. Table \ref{tab:tabVarEqTerms} provides} an overview of the result of the derivation for the terms in the above matrix equations. Note that this approach is directly applicable to $i,j>2$ as (in the absence of tides) any two-body interaction is independent of additional bodies in the system under consideration.

Since the equations of motion only contain the relative position $\mathbf{r}$, not the absolute position $\mathbf{r}_{i}$, we have the following symmetry relation: 
\begin{align}
\frac{\partial *}{\partial \mathbf{r}_{2}}=-\frac{\partial *}{\partial \mathbf{r}_{1}}=\frac{\partial *}{\partial \mathbf{r}}\label{eq:generalR1R2Partial}
\end{align}
Also, the symmetry expressed by Eq. (\ref{eq:body2Translation}), allows for the computation of the partial derivatives of $\mathbf{v}_{2}$ from the associated partials of $\dot{\mathbf{v}}_{1}$ as:
\begin{align}
\frac{\partial\dot{\mathbf{v}}_{2}}{\partial{*}}&=-\frac{M_{1}}{M_{2}}\frac{\partial\dot{\mathbf{v}}_{1}}{\partial{*}}\label{eq:v2PartialGeneral}
\end{align}
No such symmetry exists for the partial derivatives w.r.t. $\mathbf{q}_{j}$, since the potential is explicitly dependent on both $\boldsymbol{\mathcal{R}}^{\nicefrac{\mathcal{F}_{1}}{\mathcal{F}_{2}}}$ and $\boldsymbol{\mathcal{R}}^{\nicefrac{\mathcal{F}_{1}}{I}}$, but not $\boldsymbol{\mathcal{R}}^{\nicefrac{\mathcal{F}_{2}}{I}}$. {This is due to the fact that the angles $\vartheta$ and $\varphi$, representing the relative position of body 2 w.r.t. body 1, are expressed in a frame fixed to body 1. 

%
Having defined the relevant symmetry relations, we derive expressions for the partial derivatives in Eq. (\ref{eq:stateDerivativePartialMatrix}). 
The partial derivatives of $\dot{\mathbf{v}}_{1}$ w.r.t. positions $\mathbf{r}_{i}$ are obtained from Eqs. (\ref{eq:body1Translation}):
\begin{align}
\frac{\partial\dot{\mathbf{v}}_{1}}{\partial{\mathbf{r}}}&=\frac{1}{M_{1}}\boldsymbol{\mathcal{R}}^{\nicefrac{I}{\mathcal{F}_{1}}}\left(\frac{\partial^{2} V_{1\text{-}2}}{\partial \left(\mathbf{r}^{\mathcal{F}_{1}}\right)^{2}}\right)\boldsymbol{\mathcal{R}}^{\nicefrac{\mathcal{F}_{1}}{I}}\label{eq:v1DotWrtR}
\end{align}
requiring the computation of the second derivatives of the potential components w.r.t. $\mathbf{r}^{\mathcal{F}_{1}}$, which we discuss later in this section. {Note the post-multiplication with $\boldsymbol{\mathcal{R}}^{\nicefrac{\mathcal{F}_{1}}{I}}$ (in both Eq. (\ref{eq:v1DotWrtR}) and Eq. (\ref{eq:momentOnBody2WrtR})), which is due to the potential Hessian being computed w.r.t. the body-fixed position $\mathbf{r}^{\mathcal{F}_{1}}$, whereas the required partial derivative for Eq. (\ref{eq:stateDerivativePartialMatrix}) is required w.r.t. the inertial position $\mathbf{r}$.}

\begin{table}[tbp]
\small
\caption{List of equations used to evaluate terms in Eq. (\ref{eq:stateDerivativePartialMatrix}). Both the primary equations, as well as symmetry equations (in both $i$ and $j$) used to reduce computational load are given.}
\label{tab:tabVarEqTerms}
\centering
\begin{tabular}{c | c c c c}
\hline
\hline
Partial terms & Primary Eqs. & Symmetry Eqs. & Potential Partials & Angle Partials \\
\hline
$\frac{\partial\dot{\boldsymbol{\omega}}_{i}}{\partial\boldsymbol{\omega}_{j}}$ &  (\ref{eq:omegaDotOmegaPartial})  & - & - & - \\
$\frac{\partial\dot{\mathbf{v}}_{i}}{\partial\mathbf{r}_{j}}$& (\ref{eq:v1DotWrtR}) &  (\ref{eq:v2PartialGeneral}), (\ref{eq:generalR1R2Partial}) & $\frac{\partial^{2}}{\partial \left(\mathbf{r}^{\mathcal{F}_{1}}\right)^{2}}$ & -\\  
$\frac{\partial\dot{\mathbf{v}}_{i}}{\partial\mathbf{q}_{j}}$ & (\ref{eq:dvdotdq}) & (\ref{eq:v2PartialGeneral})  & $\frac{\partial}{\partial \mathbf{r}^{\mathcal{F}_{1}}\partial \mathbf{c}}$ $\frac{\partial }{\partial \mathbf{r}^{\mathcal{F}_{1}}}$ & See list in Eq. (\ref{eq:angleDerivatives}) \\
$\frac{\partial\dot{\boldsymbol{\omega}}_{i}}{\partial\mathbf{r}_{j}}$ & (\ref{eq:omegaDotMomentpartialRelation}), (\ref{eq:momentOnBody1WrtR}) & (\ref{eq:generalR1R2Partial})  & $\frac{\partial^{2}}{\partial \left(\mathbf{r}^{\mathcal{F}_{1}}\right)^{2}}$, $\frac{\partial }{\partial\mathbf{r}^{\mathcal{F}_{1}}}$ & - \\
$\frac{\partial\dot{\boldsymbol{\omega}}_{i}}{\partial\mathbf{q}_{j}}$ & (\ref{eq:omegaDotMomentpartialRelation}), (\ref{eq:momentOnBody1WrtQ}), (\ref{eq:M2WrtQj}) & - & See list in Eq. (\ref{eq:potentialDerivatives}) & See list in Eq. (\ref{eq:angleDerivatives})\\
\hline
\hline
\end{tabular}
\end{table}

The expression for the partial derivatives of $\dot{\mathbf{v}}_{1}$ w.r.t. the orientations $\mathbf{q}_{j}$ are obtained from:
\begin{align}
\frac{\partial\dot{\mathbf{v}}_{1}}{\partial{\mathbf{q}_{j}}}&=\frac{1}{M_{1}}\left(\frac{\partial \boldsymbol{\mathcal{R}}^{\nicefrac{I}{\mathcal{F}_{1}}}}{\partial \mathbf{q}_{j}}\frac{\partial V_{1\text{-}2}}{\partial\mathbf{r}^{\mathcal{F}_{1}}}+\boldsymbol{\mathcal{R}}^{\nicefrac{I}{\mathcal{F}_{1}}}\frac{\partial}{\partial \mathbf{q}_{j}}\left(\frac{\partial V_{1\text{-}2}}{\partial \mathbf{r}^{\mathcal{F}_{1}}}\right)\right)\label{eq:dvdotdq}
\end{align}
where the derivatives of $\boldsymbol{\mathcal{R}}^{\nicefrac{I}{\mathcal{F}_{1}}}$ are only non-zero for $j=1$ (\emph{i.e.,} for the partial w.r.t. the orientation of body 1). Since the potential is written in terms of $\mathbf{r}^{\mathcal{F}_{1}}$, not the state variable $\mathbf{r}$, the term $\frac{\partial}{\partial \mathbf{q}_{j}}\left(\frac{\partial V_{1\text{-}2}}{\partial \mathbf{r}^{\mathcal{F}_{1}}}\right)$ is computed using:
\begin{align}
\frac{\partial *}{\partial \mathbf{q}_{j}}&=\frac{\partial *}{\partial \mathbf{c}}\frac{\partial \mathbf{c}}{\partial\mathbf{q}_{j}}+\frac{\partial *}{\partial \left(\mathbf{r}^{\mathcal{F}_{1}}\right)}\frac{\partial \left(\boldsymbol{\mathcal{R}}^{\nicefrac{I}{\mathcal{F}_{1}}}\right)^{T}}{\partial \mathbf{q}_{j}}\mathbf{r}\label{eq:potentialQDerivative}
\end{align}
where $*=\frac{\partial V_{1\text{-}2}}{\partial \mathbf{r}^{\mathcal{F}_{1}}}$ when evaluating Eq. (\ref{eq:dvdotdq}). The cross-derivative of $V_{1\text{-}2}$ is discussed later in this section.
The partial derivatives ${\partial \mathbf{c}}/{\partial \mathbf{q}_{i}}$ are computed from ${\partial \mathbf{c}}/{\partial \mathbf{q}}$ and ${\partial \mathbf{q}}/{\partial \mathbf{q}_{i}}$, which are explicitly given in \ref{app:quatIntro} and \ref{app:cayleyKlein}, respectively.  

The derivative of $\dot{\boldsymbol{\omega}}_{i}$ w.r.t. ${\boldsymbol{\omega}}_{i}$, which follows directly from Eq. (\ref{eq:bodyiRotationalDynamics}), with $\dot{\mathbf{I}}_{i}=0$:
\begin{align}
\frac{\partial\dot{\boldsymbol{\omega}}_{i}}{\partial\boldsymbol{\omega}_{i}}&=\mathbf{I}_{i}^{-1}\left(-\frac{\partial[\boldsymbol{\omega}_{i}]_{\times}}{\partial\boldsymbol{\omega}_{i}}\mathbf{I}_{i}\boldsymbol{\omega}_{i}-[\boldsymbol{\omega}_{i}]_{\times}\mathbf{I}_{i}\right)\label{eq:omegaDotOmegaPartial}\\
\mathbf{c}\times\mathbf{b}&=[\mathbf{c}]_{\times}\mathbf{b}\label{eq:matrixProduct}
\end{align}
{where the term $[\mathbf{c}]_{\times}$ is an anti-symmetric matrix used to represent the cross-product (as this will aid later derivations), and is defined as:}
\begin{align}
[\mathbf{c}]_{\times}=\begin{pmatrix} 0 & -c_{3} & c_{2} \\ c_{3} & 0 & -c_{1} \\ -c_{2} & c_{1} & 0 \end{pmatrix}
\end{align}
{with $\mathbf{c}=[c_{1},\,\,c_{2},\,\,c_{3}]^{T}$.}

For the partial derivatives of $\dot{\boldsymbol{\omega}}_{i}$ w.r.t. the $\mathbf{r}_{j}$ and  $\mathbf{q}_{j}$, we obtain from Eq. (\ref{eq:body1RotationalKinematics}):
\begin{align}
\frac{\partial \dot{\boldsymbol{\omega}}_{i}}{\partial *}=\mathbf{I}_{i}^{-1}\frac{\partial \dot{\mathbf{M}}_{i}}{\partial *}\label{eq:omegaDotMomentpartialRelation}
\end{align}
The partial derivative of $\mathbf{M}_{1}^{\mathcal{F}_{1}}$ w.r.t. $\mathbf{r}$ are obtained from Eq. (\ref{eq:moment1Formulation}):
\begin{align}
\frac{\partial\mathbf{M}_{1}^{\mathcal{F}_{1}}}{\partial\mathbf{r}}&=\boldsymbol{\mathcal{R}}^{\nicefrac{\mathcal{F}_{1}}{I}}\times\left(\frac{\partial V_{1\text{-}2}}{\partial \mathbf{r}^{\mathcal{F}_{1}}}\right)+\mathbf{r}^{\mathcal{F}_{1}}\times\left(\frac{\partial^{2} V_{1\text{-}2}}{\partial \left(\mathbf{r}^{\mathcal{F}_{1}}\right)^{2}}\boldsymbol{\mathcal{R}}^{\nicefrac{\mathcal{F}_{1}}{I}}\right)-\frac{\partial\mathbf{M}_{2}^{\mathcal{F}_{1}}}{\partial\mathbf{r}}\label{eq:momentOnBody1WrtR}\\\
\frac{\partial\mathbf{M}_{2}^{\mathcal{F}_{1}}}{\partial\mathbf{r}}&=-\frac{\partial \boldsymbol{\hat{\mathcal{J}}}\left(V_{1\text{-}2}\right)}{\partial\mathbf{r}^{\mathcal{F}_{1}}}\boldsymbol{\mathcal{R}}^{\nicefrac{\mathcal{F}_{1}}{I}}\label{eq:momentOnBody2WrtR}
\end{align}
The final term can be computed using the same procedure as $\frac{\partial V_{1\text{-}2}}{\partial \mathbf{r}^{\mathcal{F}_{1}}}$.

For the final partial derivatives ${\partial\mathbf{M}_{i}^{\mathcal{F}_{i}}}/{\partial\mathbf{q}_{j}}$, {none of the preceding symmetry relations can be used.} From Eq. (\ref{eq:moment1Formulation}), we obtain the following for $i=1$.
\begin{align}
&\frac{\partial\mathbf{M}_{1}^{\mathcal{F}_{1}}}{\partial\mathbf{q}_{j}}=\left(\frac{\partial \boldsymbol{\mathcal{R}}^{\nicefrac{\mathcal{F}_{1}}{I}}}{\partial \mathbf{q}_{j}}\mathbf{r}\right)\times\frac{\partial V_{1\text{-}2}}{\partial\mathbf{r}^{\mathcal{F}_{1}}}+
\mathbf{r}^{\mathcal{F}_{1}}\times\left(\frac{\partial^{2} V_{1\text{-}2}}{\partial \mathbf{r}^{\mathcal{F}_{1}}\partial \mathbf{q}_{j}}\right)-\frac{\partial\mathbf{M}_{2}^{\mathcal{F}_{1}}}{\partial\mathbf{q}_{j}}\label{eq:momentOnBody1WrtQ}\\
&\frac{\partial\mathbf{M}_{2}^{\mathcal{F}_{1}}}{\partial\mathbf{q}_{j}}=-\frac{\partial \boldsymbol{\hat{\mathcal{J}}}\left(V_{1\text{-}2}\right)}{\partial\mathbf{q}_{j}}\label{eq:momentOnBody21WrtQ}
\end{align}
where again the derivatives of $\boldsymbol{\mathcal{R}}^{\nicefrac{\mathcal{F}_{1}}{I}}$ are non-zero only for $j=1$. Note that the final term in Eq. (\ref{eq:momentOnBody21WrtQ}) is computed using Eq. (\ref{eq:potentialQDerivative}). 

Finally, for the partial derivatives of $\mathbf{M}_{2}^{\mathcal{F}_{2}}$, we have from Eq. (\ref{eq:moment2FormulationInFrame1}):
\begin{align}
\frac{\partial\mathbf{M}_{2}^{\mathcal{F}_{2}}}{\partial\mathbf{r}}&=\boldsymbol{\mathcal{R}}^{\nicefrac{\mathcal{F}_{2}}{\mathcal{F}_{1}}}\frac{\partial\mathbf{M}_{2}^{\mathcal{F}_{1}}}{\partial\mathbf{r}}\label{eq:M2WrtR}\\
\frac{\partial\mathbf{M}_{2}^{\mathcal{F}_{2}}}{\partial\mathbf{q}_{j}}&=\boldsymbol{\mathcal{R}}^{\nicefrac{\mathcal{F}_{2}}{\mathcal{F}_{1}}}\left(\frac{\partial\mathbf{M}_{2}^{\mathcal{F}_{1}}}{\partial\mathbf{q}_{j}}\right)+\frac{\partial \boldsymbol{\mathcal{R}}^{\nicefrac{\mathcal{F}_{2}}{\mathcal{F}_{1}}}}{\partial \mathbf{q}_{j}}\mathbf{M}_{2}^{\mathcal{F}_{1}}\label{eq:M2WrtQj}
\end{align}

From Eqs. (\ref{eq:generalR1R2Partial})-(\ref{eq:M2WrtQj}), we can compute the terms in Eq. (\ref{eq:stateDerivativePartialMatrix}), as summarized in Table \ref{tab:tabVarEqTerms}. 
{The folowing partial derivatives of the mutual potential are needed in the formulation}:
\begin{align}
\frac{\partial V_{1\text{-}2}}{\partial \mathbf{r}^{\mathcal{F}_{1}} },\frac{\partial^{2}V_{1\text{-}2}}{\partial \left(\mathbf{r}^{\mathcal{F}_{1}}\right)^{2}},
\frac{\partial^{2}V_{1\text{-}2}}{\partial \mathbf{r}^{\mathcal{F}_{1}}\partial \mathbf{c}}, \frac{\partial \boldsymbol{\hat{\mathcal{J}}}\left(V_{1\text{-}2}\right)}{\partial\mathbf{r}^{\mathcal{F}_{1}}}, \frac{\partial \boldsymbol{\hat{\mathcal{J}}}\left(V_{1\text{-}2}\right)}{\partial\mathbf{c}}\label{eq:potentialDerivatives}
\end{align}
where each of the partial derivatives can be evaluated in a component-wise manner on $V_{l_{2},m_{2}}^{l_{1},m_{1}}$ and $\boldsymbol{\hat{\mathcal{J}}}(V_{l_{2},m_{2}}^{l_{1},m_{1}})$.

The calculation of the $\frac{\partial^{2}V}{\partial (\mathbf{r}^{\mathcal{F}_{1}})^{2}}$ can be performed using well-known techniques in space geodesy, \emph{e.g. } \cite{MontenbruckGill2000}. 
{We obtain the derivatives of $V_{l_{2},m_{2}}^{l_{1},m_{1}}$ w.r.t. the Cayley-Klein parameters $\mathbf{c}$ from Eqs. (\ref{eq:normalizedCoefficientsTransformation}) and (\ref{eq:wignerRecursive})}:
\begin{align}
\frac{\partial V_{l_{2},m_{2}}^{l_{1},m_{1}}}{\partial \mathbf{c}}&=\bar{\gamma}_{l_{2},m_{2}}^{l_{1},m_{1}}R_{1}^{l_{1}}R_{2}^{l_{2}}\frac{\bar{Y}_{l_{1}+l_{2},m_{1}+m_{2}}(\vartheta,\varphi)}{r^{l_{1}+l_{2}+1}}\frac{\partial\bar{{\mathcal{M}}}_{l_{1},m_{1},l_{2},m_{2}}^{1,2;\mathcal{F}_{1}}}{\partial\bar{\boldsymbol{\mathcal{M}}}_{l_{2},m_{2}}^{2,\mathcal{F}_{1}}}\frac{\partial\bar{\boldsymbol{\mathcal{M}}}_{l_{2},m_{2}}^{2,\mathcal{F}_{1}}}{\partial\mathbf{c}}\label{eq:singlePotentialTermpartialTheta}\\
\frac{\partial^{2}V_{l_{2},m_{2}}^{l_{1},m_{1}}}{\partial \mathbf{r}^{\mathcal{F}_{1}}\partial \mathbf{c}}&=\bar{\gamma}_{l_{2},m_{2}}^{l_{1},m_{1}}R_{1}^{l_{1}}R_{2}^{l_{2}}\frac{\partial}{\partial\mathbf{r}^{\mathcal{F}_{1}}}\left(\frac{Y_{l_{1}+l_{2},m_{1}+m_{2}}(\vartheta,\varphi)}{r^{l_{1}+l_{2}+1}}\right)\frac{\partial\bar{{\mathcal{M}}}_{l_{1},m_{1},l_{2},m_{2}}^{1,2;\mathcal{F}_{1}}}{\partial\bar{\boldsymbol{\mathcal{M}}}_{l_{2},m_{2}}^{2,\mathcal{F}_{1}}}\frac{\partial\bar{\boldsymbol{\mathcal{M}}}_{l_{2},m_{2}}^{2,\mathcal{F}_{1}}}{\partial\mathbf{c}}\\
\frac{\partial{\mathcal{\bar{M}}}_{lm}^{2,\mathcal{F}_{1}}}{\partial\mathbf{c}}&=\sum_{k=-l}^{l}\bar{\nu}_{lmk}\frac{\partial D_{mk}^{l}}{\partial\mathbf{c}}{\mathcal{\bar{M}}}_{lk}^{2,\mathcal{F}_{2}}\\
\frac{\partial D^ {l}_{mk}}{\partial\mathbf{c}}&=\sum_{p=-1}^ {1}c_{mk}^{l;p}\left(\frac{\partial D^{1}_{1,-p}}{\partial\mathbf{c}}D^{l-1}_{m-1,k+p}+D^{1}_{1,-p}\frac{\partial D^{l-1}_{m-1,k+p}}{\partial\mathbf{c}}\right),\,\,\,\,l>1\label{eq:wignerPartials}
\end{align}
where we have introduced the real column vector $\bar{\boldsymbol{\mathcal{M}}}_{l_{2},m_{2}}^{2,\mathcal{F}_{1}}=[\Re\left(\bar{{\mathcal{M}}}_{l_{2},m_{2}}^{2,\mathcal{F}_{1}}\right),\Im\left(\bar{{\mathcal{M}}}_{l_{2},m_{2}}^{2,\mathcal{F}_{1}}\right)]^{T}$.
The derivatives $\partial D^{l}_{mk}/\partial\mathbf{c}$ for $l=0,1$ are obtained directly from Eqs. (\ref{eq:ckVector}) and (\ref{eq:wignerRecursionInitialization}).

%

The derivatives of the angular momentum operators are obtained similarly:
\begin{align}
\frac{\partial \hat{\boldsymbol{\mathcal{J}}}(V_{l_{2},m_{2}}^{l_{1},m_{1}})}{\partial \mathbf{\mathbf{r}}^{\mathcal{F}_{1}}}&=
\hat{\boldsymbol{\mathcal{J}}}(\mathcal{\bar{M}}_{l_{2},m_{2}}^{2,\mathcal{F}_{1}})\bar{u}^{l_{1},m_{1}}_{l_{2},m_{2}} \frac{\partial}{\partial\mathbf{r}^{\mathcal{F}_{1}}}\left(\frac{Y_{l_{1}+l_{2},m_{1}+m_{2}}(\vartheta,\varphi)}{r^{l_{1}+l_{2}+1}}\right)\\
\frac{\partial \hat{\boldsymbol{\mathcal{J}}}(V_{l_{2},m_{2}}^{l_{1},m_{1}})}{\partial \mathbf{c}}&=\frac{\partial\hat{\boldsymbol{\mathcal{J}}}(\mathcal{\bar{M}}_{l_{2},m_{2}}^{2,\mathcal{F}_{1}})}{\partial\mathbf{c}}\bar{u}^{l_{1},m_{1}}_{l_{2},m_{2}} \frac{\bar{Y}_{l_{1}+l_{2},m_{1}+m_{2}}(\vartheta,\varphi)}{r^{l_{1}+l_{2}+1}}\\
\frac{\partial\hat{\boldsymbol{\mathcal{J}}}(\mathcal{\bar{M}}_{l_{2},m_{2}}^{2,\mathcal{F}_{1}})}{\partial\mathbf{c}}&=\sum_{k_{2}=-l_{2}}^{l_{2}}\bar{\nu}_{lmk}\mathcal{\bar{M}}_{l_{2},k_{2}}^{2,\mathcal{F}_{2}}\mathbf{K}_{l_{2}m_{2}}\frac{\partial\hat{\mathbf{D}}^{l}_{m,k}}{\partial\mathbf{c}}\label{eq:angularMomentumOperatorWrtC}
\end{align}
where $\frac{\partial\hat{\mathbf{D}}^{l}_{m,k}}{\partial\mathbf{c}}$ follows directly from Eqs. (\ref{eq:angMomOnWigner}) and (\ref{eq:wignerPartials}).

In addition to the partial derivatives of the potential terms, partial derivatives between the angle representations are required for Eqs. (\ref{eq:v1DotWrtR})-(\ref{eq:M2WrtQj}). These relations are discussed in \ref{app:angleQuaternionTransformations}:
\begingroup
\allowdisplaybreaks[3]
\begin{align}
\underbrace{\frac{\partial\mathbf{c}}{\partial\mathbf{q}}}_{\text{Eq. (\ref{eq:dCDq})}},\,\,\,\,\,\,\,\underbrace{\frac{\partial\mathbf{q}}{\partial\mathbf{q}_{j}}}_{\text{Eqs. (\ref{eq:q1WrtQ})- (\ref{eq:q2WrtQ})}},\,\,\,\,\,\,\underbrace{ \frac{\partial \boldsymbol{\mathcal{R}}^{\nicefrac{I}{\mathcal{F}_{1}}}}{\partial \mathbf{q}_{1}}}_{\text{Eqs.(\ref{eq:rotationMatrixInvariants})-(\ref{eq:R3qDependency})}} \label{eq:angleDerivatives}
\end{align}
\endgroup
finalizing the completely analytical model for the evaluation of $\dot{\boldsymbol{\Phi}}$ of the full two-body gravitational interaction.

\subsubsection{Sensitivity matrix}
\label{sec:sensitivity}

A formulation for ${\partial\dot{\mathbf{x}}}/{\partial\mathbf{p}}$ is required for the evaluation of Eq. (\ref{eq:varEqSens}). The parameter vector $\mathbf{p}$ can contain any physical parameter of the environment/system/observable. In the context of the F2BP, key parameters are the static gravity field coefficients of both bodies. Here, we provide a general formulation of ${\partial\dot{\mathbf{x}}}/{\partial\mathbf{p}}$, for the case where entries of $\mathbf{p}$ directly influence $(\bar{C},\bar{S})_{lm}^{i,\mathcal{F}_{i}}$.
By using such a formulation, the influence of model parameters on the full dynamics are accurately {and consistently represented. This will prevent errors in the state transition/sensitivity matrices from propagating into biased estimates of the gravity field parameters. This is illustrated with test cases of Phobos and KW4 in Section \ref{sec:results}.}

The following partial derivatives, along with the symmetry relation in Eq. (\ref{eq:v2PartialGeneral}), allows for an analytical evaluation of ${\partial\dot{\mathbf{x}}}/{\partial\mathbf{p}}$ {(again omitting the contribution by $\dot{\mathbf{I}}$)}:
\begin{align}
\frac{\partial\dot{\mathbf{v}}_{1}}{\partial \mathbf{p}}
&=GM_{2}\boldsymbol{\mathcal{R}}^{\nicefrac{I}{\mathcal{F}_{1}}}\frac{\partial}{\partial\mathbf{p}}\left(\frac{\partial V_{1\text{-}2}}{\partial\mathbf{r}^{\mathcal{F}_{1}}}\right)^{T}\label{eq:accelerationParamaterPartial}\\
\frac{\partial\dot{\boldsymbol{\omega}}_{i}^{\mathcal{F}_{i}}}{\partial\mathbf{p}}&=\frac{\partial\mathbf{I}_{i}^{-1}}{\partial\mathbf{p}}\left(\mathbf{I}_{i}\dot{\boldsymbol{\omega}}_{i}^{\mathcal{F}_{i}} \right)+ \mathbf{I}_{i}^{-1}\left( \left(\frac{\partial\mathbf{I}_{i}}{\partial\mathbf{p}} {\boldsymbol{\omega}}_{i}^{\mathcal{F}_{i}}\right)\times\boldsymbol{\omega}_{i}^{\mathcal{F}_{i}} +\frac{\partial\boldsymbol{M}_{i}^{\mathcal{F}_{i}}}{\partial\mathbf{p}}\right)\\
\frac{\partial\boldsymbol{M}_{1}^{\mathcal{F}_{1}}}{\partial\mathbf{p}}&=\mathbf{r}^{\mathcal{F}_{1}}\times\frac{\partial}{\partial\mathbf{p}}\left(\frac{\partial V_{1\text{-}2}}{\partial\mathbf{r}^{\mathcal{F}_{1}}}\right)^{T}+\frac{\partial\hat{\boldsymbol{\mathcal{J}}}(V_{1\text{-}2})}{\partial\mathbf{p}}\\
\frac{\partial\boldsymbol{M}_{2}^{\mathcal{F}_{2}}}{\partial\mathbf{p}}&=-\boldsymbol{\mathcal{R}}^{\nicefrac{\mathcal{F}_{2}}{\mathcal{F}_{1}}}\frac{\partial\hat{\boldsymbol{\mathcal{J}}}(V_{1\text{-}2})}{\partial\mathbf{p}}
\end{align}
The derivatives of the mutual potential and angular momentum operator require the computation of the terms $\frac{\partial}{\partial \mathbf{p}}\left(\frac{\partial V_{l_{2},m_{2}}^{l_{1},m_{1}}}{\partial \mathbf{r}^{\mathcal{F}_{1}}}\right)$ and $\frac{\partial}{\partial \mathbf{p}}\left(\hat{\boldsymbol{\mathcal{J}}}(V_{l_{2},m_{2}}^{l_{1},m_{1}})\right)$, which are obtained from:

\begin{align}
\frac{\partial *}{\partial \mathbf{p}}&=
\left(\frac{\partial *}{\partial \boldsymbol{\mathcal{M}}_{l_{2},m_{2}}^{2,\mathcal{F}_{1}}}\sum _{k_{2}=-l_{2}}^{l_{2}}\left( \frac{\partial \boldsymbol{\mathcal{M}}_{l_{2},m_{2}}^{2,\mathcal{F}_{1}}}{\partial \boldsymbol{\mathcal{M}}_{l_{2},k_{2}}^{2,\mathcal{F}_{2}}}\frac{\partial \boldsymbol{\mathcal{M}}_{l_{2},k_{2}}^{2,\mathcal{F}_{2}}}{\partial \mathbf{p}}\right)+\frac{\partial *}{\partial \boldsymbol{\mathcal{M}}_{l_{1},m_{1}}^{1,\mathcal{F}_{1}}}\frac{\partial \boldsymbol{\mathcal{M}}_{l_{1},m_{1}}^{1,\mathcal{F}_{1}}}{\partial \mathbf{p}}\right)
\end{align}
The partial derivative of the inertia tensor is computed from:
\begin{align}
\frac{\partial\mathbf{I}_{i}^{-1}}{\partial\mathbf{p}}=-\mathbf{I}_{i}^{-1}\frac{\partial\mathbf{I}_{i}}{\partial\mathbf{p}}\mathbf{I}_{i}^{-1}\label{eq:inertiaTensorParameterPartial}
\end{align}
where $\mathbf{I}$ is related to the unnormalized gravity field coefficients by:
\begin{align}
\mathbf{I}=MR^{2}\left(\begin{pmatrix}\frac{C_{20}}{3}-2C_{22} & -2S_{22} & -C_{21} \\
-2S_{22} & \frac{C_{20}}{3}+2C_{22} & -S_{21} \\
-C_{21} & -S_{21} & -\frac{2C_{20}}{3}  \end{pmatrix}+\bar{I}\mathbf{1}_{3\times 3}\right)
\end{align}
Here $\bar{I}$ denotes the body's mean moment of inertia. Note that $\bar{I}$ could be one of the entries in the vector $\mathbf{p}$.

\subsection{Numerical Implementation}
\label{sec:numericalResults}

In this section we summarize the implementation of the algorithm. Our implementation has been done in an extended version \citep{Dirkx2015c} of the Tudat\footnote{The algorithms presented here have not yet been included in the publicly available repository. An overhaul and extension of project is underway at the time of writing (September 2018), with the methods presented here to be included in a later release.} software toolkit, see \ref{app:tudat}, a generic and modular astrodynamics toolbox written in C++. 
{We focus on the implementation} of the inner loops of the function evaluation (\emph{i.e}, terms inside one or more summations). 

Summarizing, the main steps in the evaluation of the governing equations are (per time step):
\begin{enumerate}
\item Computation of Wigner D-matrices $D_{mk}^{l}$ from Eqs. (\ref{eq:wignerRecursionInitialization})-(\ref{eq:wignerDSymmetry}), from the current relative orientation of the two bodies, expressed by the Cayley-Klein parameters $a$ and $b$. These are obtained directly from the components of $\mathbf{q}_{1}$ and $\mathbf{q}_{2}$ in the state vector. The relation with Cayley-Klein parameters is given in \ref{app:cayleyKlein}.
\item Computation of $\bar{C}_{l_{2}m_{2}}^{2,\mathcal{F}_{1}}$ and $\bar{S}_{l_{2}m_{2}}^{2,\mathcal{F}_{1}}$ (for all $l_{2},m_{2}$), by inserting $D_{lm}^{k}$ into Eqs. (\ref{eq:transformedCCoefficientExplicit}) and (\ref{eq:transformedSCoefficientExplicit}).
\item Evaluation of effective spherical harmonic coefficients $ \bar{C}_{l_{1,2};m_{1,2}}$ and $\bar{S}_{l_{1,2};m_{1,2}}$ for each combination of $l_{1},m_{1},l_{2},m_{2}$ from Eqs. (\ref{eq:explicitEffectiveNormalizedCosineTerms}) and (\ref{eq:explicitEffectiveNormalizedSineTerms}).
\item Evaluation of Legendre polynomials $\bar{P}_{lm}(\sin\varphi)$, using recursive algorithm from \emph{, e.g.,} \cite{MontenbruckGill2000}, and recurring terms $\cos(m\vartheta)$, $\sin(m\vartheta)$, $(R_{1}/r)^{l_{1}}$, $(R_{2}/r)^{l_{1}}$ for $0\le l\le (l_{1,max}+l_{2,max})$, $0\le m\le (m_{1,max}+m_{2,max})$.

\item Evaluation of $\partial{V_{1-2}}/\partial\mathbf{r}^{\mathcal{F}_{1}}$ (and $\partial^{2}{V_{1\text{-}2}}/\partial\left(\mathbf{r}^{\mathcal{F}_{1}}\right)^{2}$ if propagating variational equations), which are obtained directly from Eq. (\ref{eq:mutualForcePotentialFromOneBody}) and the one-body formulations given by \emph{e.g.}, \cite{MontenbruckGill2000}.

\item Evaluation of $\boldsymbol{\hat{\mathcal{J}}}\left(\mathcal{\bar{M}}_{l_{2},m_{2}}^{2,\mathcal{F}_{1}}\right)$ from Eqs. (\ref{eq:singleMultipoleAngularMomentumOperator})-(\ref{eq:mutualForcePotentialAngMomOpFromOneBody}), as well as $\partial\boldsymbol{\hat{\mathcal{J}}}\left(\mathcal{\bar{M}}_{l_{2},m_{2}}^{2,\mathcal{F}_{1}}\right)/\partial\mathbf{c}$ from Eqs (\ref{eq:wignerPartials}) and (\ref{eq:angularMomentumOperatorWrtC}) if propagating variational equations. 
\item Evaluation of any partial derivatives w.r.t $\mathbf{p}$, as per Eqs. (\ref{eq:accelerationParamaterPartial})-(\ref{eq:inertiaTensorParameterPartial}). Computation of these terms can be done largely from calculations in step 5 and 6.

\end{enumerate}
From these steps, the equations of motion, from Eqs (\ref{eq:body1Translation})-(\ref{eq:bodyiRotationalDynamics}) can be evaluated, as well as Eqs. (\ref{eq:varEq})-(\ref{eq:varEqSens}) when propagating variational equations.

\section{Model results}
\label{sec:results}

We consider two test cases to illustrate our methodology, and to analyze the need for the use of figure-figure interactions in accurate ephemeris generation. First, we analyze the dynamics of Phobos, motivated by various recent analyses of high-accuracy tracking to future Phobos landers \citep[\emph{e.g} ][]{TuryshevEtAl2010,LeMaistreEtAl2013,DirkxEtAl2014}. Second, we apply our method to the dynamics of the KW4 double asteroid. This double asteroid has been the ubiquitous example in analyses of the figure-figure interactions.  

Our goal in this section is to ascertain the consequence of neglecting figure-figure effects during these bodies' state estimation, if high-accuracy tracking data were available. In our simulations, we introduce a difference between the truth model (used to simulate observations of the dynamics) and the estimation model (which is used to fit the observations). Our truth model includes the figure-figure interactions as in Section \ref{sec:eom}, while our estimation model does not. We consider the estimation of only the translational dynamics, and gravity field coefficients of the bodies, as is done in data analysis of current observations (see Section \ref{eq:variationalEquations}). 

{We use simulated observations of the full three-dimensional Cartesian state of the body under consideration (w.r.t. its primary). These observations cannot be realized in practice, but are most valuable in determing the sensitivity of the dynamics to various physical effects (in this case the figure-figure interactions). Specifically, it allows us to determine how much the figure-figure effects are absorbed into the estimation of other parameters when omiting these effects during the estimation. Mathematical details of this estimation approach are given by \cite{DirkxEtAl2016b}. We use noise-free observations, to ensure that any estimation error is due to the dynamical effects, not the observation uncertainty. }The dynamical model during the estimation (\emph{e.g.} without figure-figure effects) reduces to that used by, \emph{e.g.}, \cite{LaineyEtAl2004}. This dynamical model is equivalent to our formulation, without the terms where both $l_{1}>0$ and $l_{2}>0$. 


\subsection{Phobos}
\label{sec:phobosResults}
 We take the Phobos gravity field coefficients from \cite{JacobsonLainey2014}, who obtain  {$\bar{C}_{20}^{P}= -0.0473\pm 0.003$ and $\bar{C}_{22}^{P}=0.0229\pm 0.0006$}. The other Phobos gravity field coefficients are set to zero. We use the rotational model by \cite{RambauxEtAl2012}, {which was obtained from numerical integration of rotational equations of motion, including the influence of figure-figure interactions. In their rotation model, the Phobos orbit is fixed to that produced by \cite{LaineyEtAl2007}}.

In our model, the dynamics of Phobos is numerically propagated including the figure-figure interactions between Mars and Phobos up to $l_{1}=m_{1}=l_{2}=m_{2}=2$. In Fig. \ref{fig:relativeContribution}, the magnitude of the separate terms of the series expansion of the gravitational acceleration in Eq. (\ref{eq:body1Translation}) are shown over several orbits. The strongest figure-figure interactions ($l_{1}=l_{2}=m_{1}=2,m_{2}=0$) have a magnitude that is about 0.5 \% that of the weakest point-mass interaction ($l_{1}=2,l_{2}=m_{1}=m_{2}=0$). {We use the Mars gravity field by \cite{GenovaEtAl2016}, and the Mars rotation model by \cite{KonoplivEtAl2006}.}

For geophysical analysis of Phobos, the $\bar{C}_{20}^{P}$ and $\bar{C}_{22}^{P}$ coefficients are of prime interest. Results of a consider covariance analysis by \cite{DirkxEtAl2014} have shown that $\bar{C}_{22}^{P}$ could be determined to at least $10^{-9}$ using a laser ranging system on a Phobos lander, operating over a period of 5 years (and close to $10^{-7}$ after only 1 year). For $\bar{C}_{20}$, they obtain errors of $10^{-4}${ (relative error 0.2\%) }and $10^{-7}${ (relative error 0.0002\%) }after 1 and 5 years, respectively.{ Note that they only considered  $\bar{C}_{20}$ and $\bar{C}_{22}$ for the Phobos gravity field.} Here, the influence of omitting the figure-figure interactions during the estimation of these coefficients is quantified. Ideal observations of Phobos' position are simulated, with all interactions up to $l_{1}=l_{2}=m_{1}=m_{2}=2$. Then, these simulated data are used to recover the initial state of Phobos, as well as its full degree two gravity field, using a dynamical model without any figure-figure interactions.  

The results of the estimation are shown in Fig. \ref{fig:phobosCoefficientEstimate}, where the errors in $\bar{C}^{P}_{2m}$ and $\bar{S}^{P}_{2m}$ are shown as a function of the duration of the simulations. 
The errors in  $\bar{C}_{20}^{P}$ and $\bar{C}_{22}^{P}$ due to neglecting figure-figure interactions in the estimation model are, at{ $2\cdot10^{-4}$ (relative error 0.42 \%) and $1.5\cdot10^{-5}$ (relative error 0.065 \%)} respectively, much larger than the uncertainties obtained by \cite{DirkxEtAl2014}.  This unambiguously shows that precision tracking of a Phobos lander will require the use of the figure-figure interactions in both the dynamical modelling and estimation of orbit/physical parameters. Also, it shows that for the analysis of existing data of Phobos, neglecting figure-figure interactions is an acceptable assumption, as the errors we obtain are smaller by more than an order of magnitude than the formal uncertainties of these parameters reported by \cite{JacobsonLainey2014} (shown in Fig. \ref{fig:phobosCoefficientEstimate} as dashed lines). 


\begin{figure}[tbp!]
\centering
\includegraphics[width=0.8\textwidth]{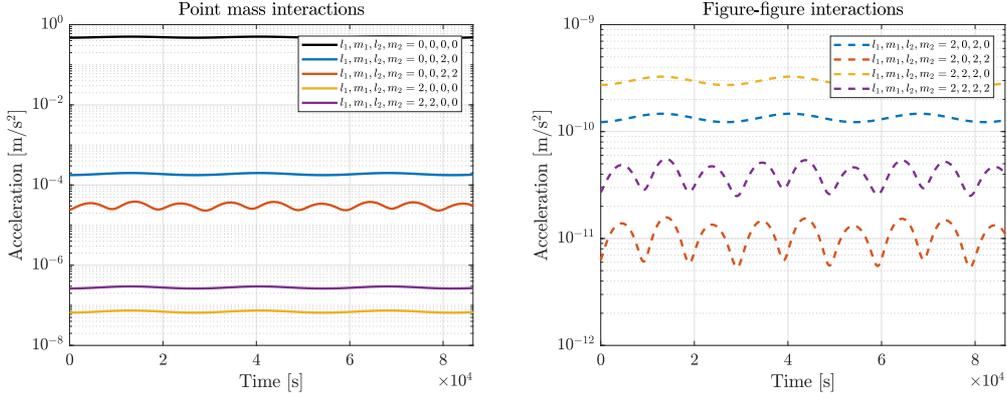}
\caption{Full two-body acceleration components acting on Phobos (index 1) due to its interaction with Mars (index 2), up to degree and order 2.}
\label{fig:relativeContribution}
\end{figure}

\begin{figure}[tbp!]
\centering
\includegraphics[width=0.8\textwidth]{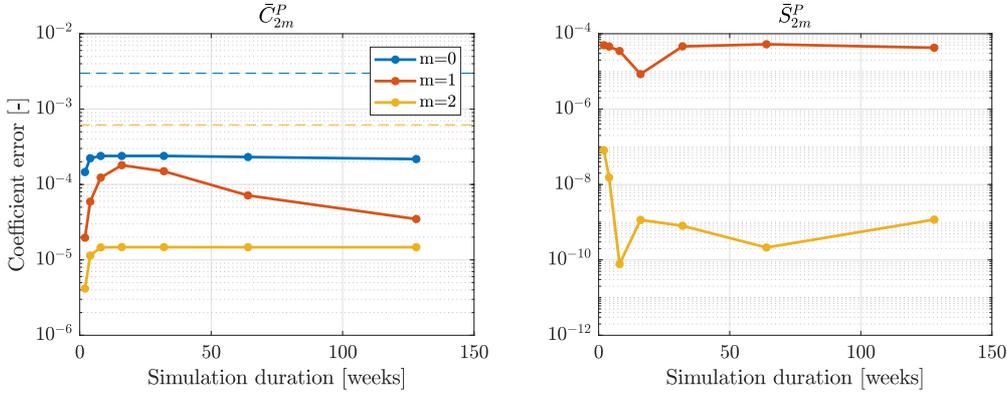}
\caption{Estimation error of degree 2 gravity field coefficients of Phobos. Observations simulated including figure-figure effects up to $l_{1}=l_{2}=2$. Estimation done without figure-figure effects, $l_{1,max}=l_{2,max}=2$. Dashed lines indicate uncertainties in $\bar{C}_{20}$ and $\bar{C}_{22}$ given by \cite{JacobsonLainey2014}.}
\label{fig:phobosCoefficientEstimate}
\end{figure}


The figure-figure interactions up to degree and order $l_{max}$ result in figure-figure interactions proportional the spherical harmonic basis-functions up to $Y_{2l_{max},2l_{max}}(\phi,\vartheta)$, see Eq. (\ref{eq:complexMutualForcePotential}). These interactions are used in the truth model, but not the estimation model, where terms up to only $Y_{l_{max},l_{max}}(\phi,\vartheta)$ are included, see Eq. (\ref{eq:singleBodyPotentialComplex}). To ascertain whether the influence of figure-figure interactions could be absorbed by estimation of terms of degree $>l_{max}$, we reran our simulations, estimating the gravity field of Phobos up to degree and order 4, instead of 2. 

Unfortunately, the resulting estimation problem becomes ill-conditioned, preventing results from being obtained. This indicates that the dynamics of Phobos does not contain the required information to independently estimate its full gravity field up to degree {four. This is a general property for satellites in a (near-)circular and (near-)equatiorial orbit, as the contribution of the degree two and degree four coefficients cannot be distinguished in the dynamics: both show the same temporal signature (with different magnitudes) in the observations, leading to ill-posedness in the estimation. The approach taken by \emph{e.g.} \cite{YoderEtAl2003} is to estimate 'lumped' gravity field coefficients, essentially acknowledging that an estimation of the $\bar{C}_{2,0}$ term includes contributions from $\bar{C}_{4,0}$, $\bar{C}_{6,0}$, \emph{etc}. As a consequence of this degeneracy, estimating higher-order gravity field coefficients will require tracking of a spacecraft around/near Phobos, as it cannot be fully estimated from Phobos' orbital dynamics alone.  
}
\begin{figure}[tbp!]
\centering
\includegraphics[width=0.87\textwidth]{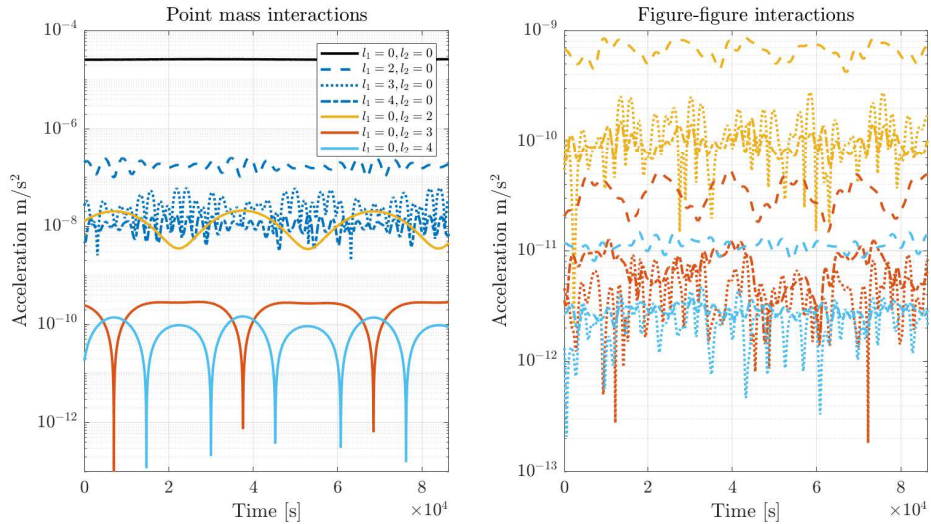}
\caption{Full two-body acceleration components acting on the body KW4-Beta, values summed over both $m_{1}$ and $m_{2}$  in Eq. \ref{eq:body1Translation}. Line style indicates value of $l_{1}$ (degree of Alfa's gravity field), line color indicates value of $l_{2}$ (degree of Beta's gravity field).}
\label{fig:kw4AccFigure}
\end{figure}

\begin{figure}[tbp!]
\centering
\includegraphics[width=0.87\textwidth]{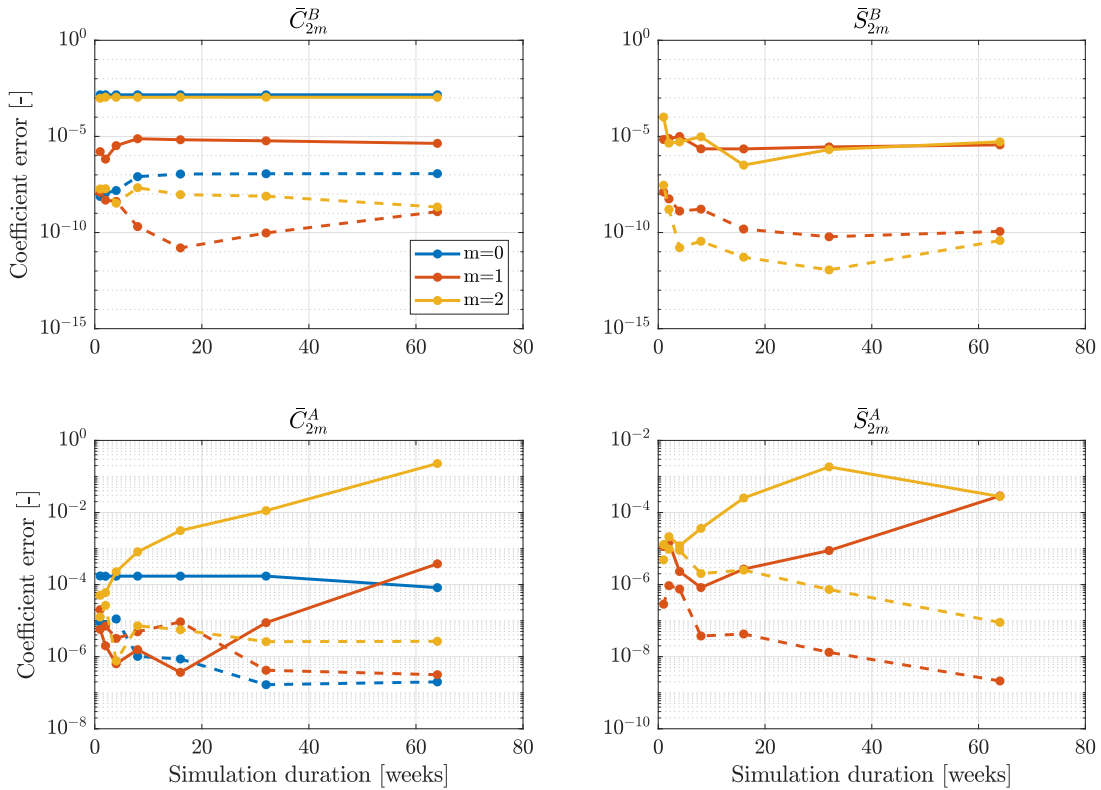}
\caption{Estimation error of degree 2 gravity field coefficients of Alfa and Beta. Observations simulated including figure-figure effects up to $l_{1}=l_{2}=2$. Continuous line: estimation without figure-figure effects, $l_{1,max}=l_{2,max}=2$. Dashed line: estimation without figure-figure effects, $l_{1,max}=l_{2,max}=4$. Coefficients of two bodies are estimated in separate simulations.}
\label{fig:kw4EstimationFigure}
\end{figure}

This section shows that future precision tracking of Phobos, in particular in the case of landers, must take into account the figure-figure interactions up to at least the 2$^{nd}$ degree of Phobos' and Mars' gravity field. This article provides the required algorithms to achieve this. The values of the gravity field coefficients encode information concerning the mass distribution inside Phobos, provide insight into the presence or absence of voids and ices in the interior, and possible lateral density variations. Failure to include these effects into the estimation model can lead to erroneous interpretation of Phobos' interior structure \citep{LeMaistreEtAl2019}. 

\subsection{KW4}

The KW4 binary asteroid system, which consists of two bodies termed Alfa and Beta, has been the 'standard' test case for the influence of figure-figure interactions. The shapes of the two asteroids were determined by \cite{OstroEtAl2006} using radar imaging. Combined with determination of the mass of the two bodies, and the assumption of homogeneous interior mass distribution, this provides a model for both bodies' gravity field, facilitating the analysis of their dynamics. The full two-body dynamics of the system was studied in detail by \cite{FahnestockScheeres2008}. The dynamics of this system was later used as a test case for models of the F2BP by \citep[\emph{e.g.}, ][]{BoueLaskar2009, CompereLemaitre2014, HouEtAl2017,Boue2017,Hou2018}.

We use the spherical harmonic coefficients of the two bodies up to degree 4, as provided by \cite{CompereLemaitre2014}. The body Beta (secondary) is propagated w.r.t. Alfa (primary), using the unexcited initial conditions given by \cite{FahnestockScheeres2008}. We plot the resulting accelerations during several orbits in Fig. \ref{fig:kw4AccFigure}. In this plot,  accelerations have been summed over both $m_{1}$ and $m_{2}$ in Eq. (\ref{eq:body1Translation}). This figure shows the significant influence that the figure-figure interactions have, with an acceleration magnitude of the strongest figure-figure interaction (total $l_{1}=l_{2}=2$ terms) at about $4\cdot10^{-5}$ that of the primary ($l_{1}=l_{2}=0$ term). In fact, the combined $l_{1}=l_{2}=2$ terms are only about one order of magnitude weaker than the combined ($l_{1}=0, l_{2}=2$) terms. This exceptionally strong influence of figure-figure interactions is a consequence of the close orbit of the two bodies ($R_{1}/a\approx 0.26$, $R_{2}/a\approx 0.09$, with $a$ the semi-major axis of the mutual orbit), and the highly irregular gravity fields of the two bodies.


Using the same methodology as for Phobos, we perform two analyses for KW4, in which we estimate only the gravity field coefficients of either Alfa \emph{or} Beta, respectively. Estimating the coefficients of both bodies simultaneously results in an ill-conditioned problem, indicating that the relative motion of the bodies alone cannot be used to constrain both bodies' gravity field coefficient{s. The underlying reason is similar to that for the ill-posedness encountered in the Phobos analysis: the contributions of, for instance, both bodies' $\bar{C}_{2,0}$ coefficients cannot be independently determined from observations of the bodies' relative dynamics. Determining both bodies' gravity fields simultaneously will require spacecraft orbit/flyby observations, which would improve the conditioning of the problem. }Such a mission design is outside of the scope of this article, and we limit ourselves to the relative motion of the two bodies.

Therefore, the results obtained here are not a direct indication of the attainable gravity field estimation accuracy during data analysis when omitting figure-figure effects, unlike the results for Phobos in Section \ref{sec:phobosResults}. Instead, our results for KW4 are indicative of the relevance of figure-figure interactions, when attempting to infer the interior structure of the bodies from their relative motion.


With $l_{max}=2$ for both the simulation and estimation model, the resulting errors in the gravity field coefficients of Alfa and Beta (which are obtained in separate estimations) are shown in Fig. \ref{fig:kw4EstimationFigure} with solid lines. As was the case for Phobos, significant errors in the estimated degree-2 gravity-field coefficients are obtained when omitting the figure-figure effects during estimation. Unlike the Phobos simulations, however, the magnitude of the errors does not always converge to a constant value for sufficiently long observation times. 

The analysis where simulation of observations was done up to $l_{max}=2$ (with figure-figure interactions), with an estimation model up to $l_{max}$=4 (without figure-figure interactions) has also been performed. Unlike the case of Phobos, where the estimation problem failed to converge in this situation, the KW4 simulations consistently produce results. The errors in the degree two coefficients of both bodies are shown in Fig. \ref{fig:kw4EstimationFigure} with dashed lines. Comparing to the case where $l_{max}=2$ during both observation simulation and estimation, there is a decrease of several orders of magnitude in the errors of the degree 2 gravity field coefficients. Moreover, the values of the estimation error show a much more stable behavior for long simulation times. Errors in the estimated degree-3 gravity-field coefficients are limited, at $10^{-5}$ at most, and $<10^{-10}$ for many coefficients. Errors in degree 4 coefficients are significant, however, in particular for Beta, where the absolute errors in $\bar{C}_{40}$ and $\bar{C}_{42}$ reach values of 0.075 and 0.012, respectively. For Alfa, these errors are at the level of $10^{-3}$ and $2\cdot10^{-4}$. {Note that Fig. \ref{fig:kw4EstimationFigure} only shows the errors in degree 2 gravity field coefficients, not in the coefficients of degree 3 or 4}.

These results indicate that the signature of figure-figure interactions at degree 2 can be partly absorbed by the estimation of gravity field coefficients of degree 4, when omitting figure-figure effects during estimation. In this situation, the degree 4 coefficients are used for a similar purpose as empirical accelerations, which are typically used in spacecraft orbit determination \citep{MontenbruckGill2000}: to absorb part of the dynamical mismodelling, and reduce the degree to which these model errors impact the estimation quality.

The impact that the estimation of higher-degree coefficients has on the attainable estimation accuracy will be strongly dependent on the system under consideration, though. Also, applying this method requires the estimation of a substantial number of extra parameters, potentially weakening the stability of the numerical solution. We stress that, even in cases where this approach could be robustly used to estimate degree-2 gravity-field coefficients, the resulting degree-4 coefficients will not have a physical meaning, {as they have absorbed part of the signature of the figure-figure effects}. When using our methodology outlined in this article for the estimation, the coefficients will not absorb dynamical model errors, and the estimated degree-4 coefficients would be representative of the bodies' actual mass distribution.

In summary, our results indicate that the determination of the gravity fields of Alfa and Beta based (in part) on the observed mutual dynamics of the two bodies should take into account figure-figure interactions. Failure to do so results in substantial errors in the degree-2 or degree-4 coefficients, depending on the degree to which the gravity fields are expanded. The attainable accuracy of gravity field coefficient estimation is strongly dependent on tracking data types/uncertainties, lander/orbiter mission geometry, \emph{etc.} For a specific mission design, our methodology can be used to quickly assess which effects should and should not be considered during the data analysis.


\section{Conclusions}
\label{sec:conclusions}
We have derived in detail the equations that relate the full two-body gravitational interaction, including all figure-figure effects, to the typical one-body gravity field representation in terms of (fully-normalized) spherical harmonic coefficients. Our derivation provides the explicit link between the elegant representations found in literature 
and the need for a {practical} implementation in existing state propagation and estimation software, in particular for applications in solar system dynamics and orbit determination. 

{Our framework enables robust modelling of the full gravitational interactions in both simulation studies and data analysis. The approach will be crucial for future tracking techniques, such as ILR and SBI, in which (sub-)mm relative model accuracy between the positions of celestial bodies is required over a period of years.}  In addition to capturing the full dynamics, the formulation allows the rotational dynamics of a celestial body to be estimated in a manner that prevents the estimation of a broad libration spectrum, and mitigates problems associated with decoupled rotational-translational dynamics in the estimation model.

We provide explicit formulations for transformed spherical harmonic coefficients, {given by} Eqs. (\ref{eq:EfunctionDecomposition})-(\ref{eq:transformedSCoefficientExplicit}), the effective coefficients for use in the one-body potential formulation in Eqs. (\ref{eq:explicitEffectiveNormalizedCosineTerms})-(\ref{eq:explicitEffectiveNormalizedSineTerms}), and the explicit transformation from two-body to one-body potential terms in Eqs. (\ref{eq:maximumEffectiveDegree})-(\ref{eq:directSComponentTransformation}) and (\ref{eq:mutualForcePotentialFromOneBody}). Associated equations for unnormalized spherical harmonic coefficients are given in \ref{app:geodesyNormalizedFormulation}. These formulations are crucial to transparently implement the F2BP in existing tracking data analysis suites.

To be able to apply our model to state estimation of natural bodies, we have derived an analytical formulation of the variational equations in Section \ref{eq:variationalEquations}. The resulting differential equations for the state transition matrix $\Phi(t,t_{0})$ and sensitivity matrix $S(t)$ allow the direct use of the models in existing state propagation and estimation software, and automatically capture any dependency of the mutual dynamics on either body's state, gravity field coefficient, or other physical parameter. The use of analytical partial derivatives is crucial to prevent an excessive computation load. 
The required equations are summarized in Table \ref{tab:tabVarEqTerms}, and have not been found comprehensively in literature for any representation of the F2BP.

The relevance of our method is predicated on the assumption that figure-figure interactions will be relevant for data analysis of (future) planetary missions. We have analyzed the impact of ignoring figure-figure interactions during state and gravity-field estimation  (as is typical standard practice) for both Phobos and the KW4 binary asteroid. For Phobos, relative errors in the estimated values of the $\bar{C}_{20}$ and  $\bar{C}_{22}$ coefficients are at{ $0.42 \%$ and $0.065\%$}, respectively. These values are below the current uncertainty in Phobos' gravity field, validating the omission of figure-figure effects in previous studies. It was shown by \cite{DirkxEtAl2014} that accurate laser tracking of a Phobos lander will allow these coefficients to be estimated to much greater accuracy, requiring the figure-figure effects to be included in the estimation model. The same will be true for Doppler tracking of a Phobos lander \citep{LeMaistreEtAl2013}. A similar analysis of the mutual dynamics of the KW4 binary asteroid system was performed, again indicating significant errors in the gravity field coefficient estimation when omitting figure-figure effects. The errors in the degree two coefficients are reduced significantly, to the range $10^{-9}$-$10^{-6}$ when increasing the degree to which gravity field coefficients are estimated from 2 to 4, while still omitting figure-figure interactions in the estimation. However, this is at the expense of significant errors in the degree 4 gravity field coefficients.

Each of the existing formulations of the F2BP, for instance in terms of mass multipoles \citep{CompereLemaitre2014}, inertia integrals \citep{HouEtAl2016} and polyhedrons \citep{FahnestockScheeres2008}, has its specific (dis)advantages. Our formulation for the F2BP in terms of spherical harmonic coefficients is purposefully designed for the use in space-mission tracking data analysis, maximizing compatibility with currently typical approaches to state and gravity field parameter estimation.  

\section*{Acknowledgements}

We thank Xiyun Hou and Nicolas Rambaux for carefully reading our manuscript, and providing key feedback and insight that improved the clarity and completeness of the article.

\begin{appendix}

\section{Relations Between Rotation Representations}
\label{app:angleQuaternionTransformations}

In this appendix, we discuss the relation between the angle representations we use: 3-1-3 Euler angles, rotation matrices and quaternions, discussed in greater detail by \cite{Diebel2006}. Additionally, we derive explicit formulations for the partial derivatives between these representations that are needed in the evaluation of the full two-body gravitational problem.

\subsection{Quaternions}
\label{app:quatIntro}
We nominally use quaternions to represent the orientation of celestial bodies. Quaternions provide a singularity-free representation, while ensuring that the minimum number of required variables are propagated \citep{Hughes2004}. Moreover, propagating the quaternions was shown by \cite{Fukushima2008} to be the optimal choice (of the broad range of options that they considered) in terms of numerical integration errors. The quaternion has four entries, denoted as $q_{0}$, $q_{1}$, $q_{2}$ and $q_{3}$, which can be directly related to an angle-axis transformation of an angle $\theta$ about a unit-axis $\hat{\mathbf{n}}$ through:
\begin{align}
\mathbf{q}&=\left(\cos\frac{\theta}{2},\hat{\mathbf{n}}\sin\frac{\theta}{2}\right)\\
&=\left(s,\mathbf{v}\right)\label{eq:quaternionSplitting}
\end{align}
where the first term denotes the $q_{0}$ entry, and the subsequent vector the $q_{1}$, $q_{2}$ and $q_{3}$ terms. We again use $\mathbf{q}$ for the vector representation of the quaternion and $q$ for the operator representation. For a quaternion representing a rotation, the norm is exactly 1, so:
\begin{align}
|\mathbf{q}|=\sqrt{q_{0}^{2}+q_{1}^{2}+q_{2}^{2}+q_{3}^{2}}=1\label{eq:quaternionNormalization}
\end{align}
Consequently, the first and second derivative of a quaternion w.r.t. to any parameter $\alpha$ must satisfy:
\begin{align}
&\sum_{i}\mathbf{q}_{i}\frac{\partial \mathbf{q}_{i}}{\partial\alpha}=0\\
&\sum_{i}\left(\left(\frac{\partial \mathbf{q}_{i}}{\partial\alpha}\right)^{2}+\mathbf{q}_{i}\frac{\partial^{2} \mathbf{q}_{i}}{\partial\alpha^{2}}\right)=0
\end{align}
{Defining a new quaternion operator as:}
\begin{align}
q_{AB}=q_{A}^{-1}q_{B}
\end{align}
{and denoting a quaternion as per Eq. (\ref{eq:quaternionSplitting}), we obtain the following:}
\begin{align}
\frac{\partial \mathbf{q}_{AB}}{\partial \mathbf{q}_{A}}&=-s_{B}\mathbf{I}_{4\times 4}+\left( \begin{array}{c c} 2s_{B}&  \mathbf{v}_{B}^{T} \\ 
\mathbf{v}_{B} & -[\mathbf{v}_{B}]_{\times} \end{array} \right)\label{eq:q1WrtQ}\\
\frac{\partial \mathbf{q}_{AB}}{\partial \mathbf{q}_{B}}&= s_{A}\mathbf{I}_{4\times 4}+\left( \begin{array}{c c} 0 &  \mathbf{v}_{A}^{T} \\
-\mathbf{v}_{A} &[\mathbf{v}_{A}]_{\times} \end{array} \right)\label{eq:q2WrtQ}
\end{align}
{{using the notation of Eqs. (\ref{eq:matrixProduct}) and (\ref{eq:quaternionSplitting})}. This allows partials of a composite quaternion to be determined from its constituents' partials.}

\subsection{Cayley-Klein Parameters}
\label{app:cayleyKlein}

Cayley-klein paramters are directly related to quaternions by the following:
\begin{align}
a=q_{0}-iq_{3}\\
b=q_{2}-iq_{1}
\end{align}
where we use the same sign-convention as \cite{Boue2017}. From here, the derivatives w.r.t. $\mathbf{q}$ are directly obtained from:
\begin{align}
\frac{\partial\mathbf{c}}{\partial\mathbf{q}}=\begin{pmatrix}1 & 0 & 0 & 0 \\
0 & 0 & 0 & -1\\
0 & 0 & 1 & 0\\
0 & -1 & 0 & 0 \end{pmatrix}\label{eq:dCDq}
\end{align}

\subsection{Rotation Matrix Partial Derivatives}
\label{eq:rotMatPartDer}
The evaluation of the rotational equations of motion require the determination of the terms ${\partial R_{ij}}/{\partial \alpha}$ for a number of quantities $\alpha$ (where $R_{ij}$ denotes the entry of the rotation matrix $\mathbf{R}$). However, as was discussed for quaternions in \ref{app:quatIntro}, the entries $R_{ij}$ are not independently chosen. In particular, the columns $\mathbf{R}_{i}$ must satisfy the following six relations  \citep[\emph{e.g., }][]{Maciejewski1995}:
\begin{align}
\mathbf{R}_{i}\cdot\mathbf{R}_{j}=2\delta_{ij}\,\,\,\,(i=1..3, j=1..3, j\ge i)\label{eq:rotationMatrixInvariants0}
\end{align}
Consequently, the derivatives of the rotation matrices w.r.t. some quantity $\alpha$ must obey:
\begin{align}
\frac{\partial\mathbf{R}_{i}}{\partial\alpha}{\cdot\mathbf{R}_{j}}+\mathbf{R}_{i}\cdot\frac{\partial\mathbf{R}_{j}}{\partial\alpha}=0\,\,\,\,(i=1..3, j=1..3, j\ge i)\label{eq:rotationMatrixInvariants}
\end{align}
Here we choose the terms $R_{13}$, $R_{23}$ and $R_{31}$ to be independent\footnote{Note that if, from the 9 components $R_{ij}$, the three components with equal $i$ or equal $j$ are chosen to be independent, the solution becomes singular.}, which we collectively term $\mathbf{P}$. The vector of the remaining six dependent entries is denoted $\mathbf{R}'$. The vector of dependent partial derivatives, denoted $\frac{\partial\mathbf{R}'}{\partial\alpha}$, is now obtained from the solution of:
\begin{align}
\mathbf{A}\left(\frac{\partial\mathbf{R}'}{\partial\alpha}\right) = \mathbf{b}\label{eq:generalMatrixEqRDerivtaive}
\end{align}
where the matrix $\mathbf{A}$ is expressed in terms of $\mathbf{P}$ and $\mathbf{R}'$, and $\mathbf{b}$ is a function of $\mathbf{R}$ and $\frac{\partial\mathbf{P}}{\partial\alpha}$. The entries of the matrix $\mathbf{A}$ represent and the vector $\mathbf{b}$ are directly obtained from Eq.  (\ref{eq:rotationMatrixInvariants}) for each of the six equations in Eq. (\ref{eq:rotationMatrixInvariants}). The above can be used to obtain $\partial R_{ij}/\partial q_{k}$ from the corresponding derivatives of $\mathbf{P}$. 

For completeness, we give the explicit formulations of the three independent entries in terms of $\mathbf{q}$:
\begin{align}
R_{13}=2(q_{1}q_{3}-q_{0}q_{2})\label{eq:R1qDependency}\\
R_{23}=2(q_{0}q_{1}+q_{2}q_{3})\\
R_{31}=2(q_{0}q_{2}+q_{1}q_{3})\label{eq:R3qDependency}
\end{align}

\section{Formulation in Unnormalized Spherical Harmonic Coefficients}
\label{app:geodesyNormalizedFormulation}
The equations derived in Sections \ref{sec:potential} and \ref{sec:dynamicalEquations} are written in terms of fully normalized spherical harmonic coefficients $\bar{C}_{lm}$ and $\bar{S}_{lm}$. As both the unnormalized and normalized coefficients are used in planetary science/space geodesy, we summarize the relevant equations for unnormalized coefficients here.

\subsection{Mutual Force Potential Term}
{The one-body potential using the unnormalized coefficients ${C}_{lm}$ and ${S}_{lm}$ and Legendre polynomials is given by Eq. (\ref{eq:singleBodyPotentialReal}). A single term $V_{l_{2},m_{2}}^{l_{1},m_{1}}$ of $V_{1-2}$ in Eq. (\ref{eq:realMutualForcePotentialNormalizedComponent}) then becomes}, using Eqs. (\ref{eq:fullNormalization}), (\ref{eq:fullNormalizationCoefficient}) and (\ref{eq:geodesyNormalizedLegendre}):
\begingroup
\allowdisplaybreaks[3]
\begin{align}
&V_{l_{2},m_{2}}^{l_{1},m_{1}}(\mathbf{r},\boldsymbol{\mathcal{R}}^{\nicefrac{\mathcal{F}_{1}}{\mathcal{F}_{2}}})={\gamma}^{l_{1},m_{1}}_{l_{2},m_{2}}\mathcal{{M}}_{l_{1},m_{1},l_{2},m_{2}}^{1,2;\mathcal{F}_{1}}\bigg(\cos(|m_{1}+m_{2}|\vartheta)+...\nonumber\\
&\hspace{1.0cm}...+i\left(s_{m_{1}+m_{2}}\sin(|m_{1}+m_{2}|\vartheta)\right)\bigg)\frac{{P}_{l_{1}+l_{2},|m_{1}+m_{2}|}(\sin\varphi)}{r}\left(\frac{R_{1}}{r}\right)^{l_{1}}\left(\frac{R_{2}}{r}\right)^{l_{2}}\\
&\gamma^{l_{1},m_{1}}_{l_{2},m_{2}}=(\text{-}1)^{l_{1}}\frac{(l_{1}+l_{2}-m_{1}-m_{2})!}{(l_{1}-m_{1})!(l_{2}-m_{2})!}\Sigma_{l_{1}+l_{2},m_{1}+m_{2}}\\
&\Sigma_{lm}=\begin{cases} 1 & m\ge 0\\ (\text{-}1)^{m}\frac{(l+m)!}{(l-m)!} & m < 0 \end{cases}\\
&\mathcal{{M}}_{l_{1},m_{1},l_{2},m_{2}}^{1,2;\mathcal{F}_{1}}=\mathcal{{M}}_{l_{1},m_{1}}^{1,\mathcal{F}_{1}}\mathcal{{M}}_{l_{2},m_{2}}^{2,\mathcal{F}_{1}}
\end{align}
\endgroup
which differs from its normalized counterpart only in the fact that $\gamma$ and ${\mathcal{M}}$ are used instead of $\bar{\gamma}$ and $\bar{\mathcal{M}}$.

\subsection{Gravity Field Coefficient Transformations}

The explicit formation of the transformed unnormalized coefficients are obtained from Eqs. (\ref{eq:semiNormalization}) and (\ref{eq:normalizedCoefficientsTransformation}). The result for normalized coefficients was given in Eqs.  (\ref{eq:transformedCCoefficientExplicit}) and (\ref{eq:transformedSCoefficientExplicit}). The corresponding equations for unnormalized coefficients are obtained by $\bar{C}\rightarrow C$,  $\bar{S}\rightarrow S$ and $\bar{\nu}_{lmk}\rightarrow{\nu}_{lmk}$, with:
\begin{align}
\nu_{lmk}&=(-1)^{m+k}\sqrt{\frac{(l-m)!(l+k)!}{(l+m)!(l-k)!}}
\end{align}

The expressions for the normalized equivalent of Eqs. (\ref{eq:directCComponentTransformation}) and (\ref{eq:directSComponentTransformation}) are then obtained directly from Eqs. (\ref{eq:normalizedSingleBodyPotentialReal}) and (\ref{eq:realMutualForcePotentialNormalizedComponent}). The resulting equations are obtained by replacing both the coefficients $C$, $S$, $\mathcal{C}$, $\mathcal{S}$ and the coefficient $\gamma$ by their normalized coefficients (denoted by the straight overbar). 

These equations enable the computation of  the mutual force potential terms using subroutines for the one-body normalized spherical harmonic potential in Eq. (\ref{eq:geodesyNormalizedLegendre}). This is achieved by applying the scaling rules in Eqs. (\ref{eq:maximumEffectiveDegree})-(\ref{eq:directSComponentTransformation}), and replacing the effective one-body coefficients with the normalized counterparts from Eqs. (\ref{eq:explicitEffectiveNormalizedCosineTerms}) and (\ref{eq:explicitEffectiveNormalizedSineTerms}).

\section{Tudat Software}
\label{app:tudat}

{The Tudat software, which was used to generate the results presented in this article, is a multi-purpose, modular, astrodynamics software suite written in C++, developed by staff and students of the Asrodynamics \& Space Missions group of Delft University of Technology. The project was started in 2010, with the current architecture having been set up in 2015. Aspects of the software are discussed by \cite{Dirkx2015c}, with detailed feature documentation (including installation guide and tutorials) at \url{tudat.tudelft.nl}, and code documentation at \url{doxygen.tudat.tudelft.nl}. The code is licensed under the BSD 3-Clause "New" or "Revised" License, and freely available from \url{github.com/tudat}. All Tudat functionality is tested in an associated (unit) test. All tests can be rerun at will by users to verify the integrity of the code after installation and/or modification.}

{In this appendix, we recall the main features of the Tudat software suite, limiting ourselves to orbit propagation and estimation functionality (omitting various other features related to, \emph{e.g}, mission design).}

\subsection{State Propagation}

{A strong design driver of the Tudat software is modularity: the software architecture makes no \emph{a priori} assumptions on bodies being considered, or on models being used for physical properties of bodies, accelerations, \emph{etc.} Instead, users are free to set up the simulation according to their own needs, choosing from a broad variety of models. To make implementation straightforward, various default models are provided (which can be overridden at will), for instance for solar system ephemerides, gravity fields and rotational models. Tudat makes no distinction between natural or artificial bodies, this distinction is introduced purely by the properties assigned to a given body.}

{Tudat is applicable to a broad range of topics, and has been used for, among others, simulations of solar system dynamics \citep{DirkxEtAl2018}, interplanetary trajectory design \citep{Musegaas2013}, planetary system dynamics \citep{KumarEtAl2015, DirkxEtAl2016b}, space debris impact predictions \citep{RonseMooij2014} and  atmospheric re-entry \citep{DirkxMooij2014}. }

{Setting up a numerical state propagation in Tudat consists of defining the following models:
\begin{itemize}
\item Environment models: all properties of bodies (ephemerides, gravity fields, shapes, atmospheres, \emph{etc.}) are stored in the environment. In the Tudat architecture, this includes properties of artificial bodies, such as engine models.
\item Dynamical model type(s) to be propagated. Tudat is currently capable of numerically propagating a body's translational state, rotational state and mass (a custom state derivative model is also provided to provide flexibility for other types of dynamics). A combination of any or all of these types of dynamics, for any number of bodies may be provided. The type of dynamics need not be the same for each body, so that the translational state and mass of a spacecraft, and translational state and rotational state of a planet, may be propagated. The dynamics may be propagated hierarchically (\emph{e.g.} spacecraft w.r.t. Moon, Moon w.r.t. Earth and Earth w.r.t. Barycenter) in a concurrent fashion. 
\item Dynamical model formulation. Depending on the dynamics type, multiple formulations for the governing equations may be used, such as Cartesian, Keplerian and Modified Equinoctial Elements for translational dynamics, and quaternions or modified Rodrigues parameters, in combination with angular velocities, for rotational dynamics. States may be propagated in a single-arc or multi-arc fashion, as well as a combination of the two, as described by \citep{DirkxEtAl2018}.
\item State derivative models. Depending on the types of dynamics being propagated, models for, \emph{e.g.}, torques and accelerations must be provided. In Tudat, these models are defined by their type (\emph{e.g.} aerodynamic, spherical harmonic gravity, third-body point-mass gravity, radiation pressure), the body exerting and body undergoing the torque/acceleration and, if needed, additional information (such as maximum degree/order for a spherical harmonic acceleration).
\item Numerical integrator settings: Tudat provides implementations of fixed- and variable time-step multi-stage methods (Runge-Kutta-Fehlberg), multi-step methods (Adams-Bashforth-Moulton) and extrapolation methods (Bulirsch-Stoer). 
\item Output settings. By default, Tudat outputs only the numerically propagated states, but a wide array of additional variables may be saved in addition. These include (but are not limited to), acceleration/torque terms, local atmospheric conditions, relative orientations of bodies, \emph{etc.} Such settings may also be used to define termination conditions for a numerical propagation, for instance when two bodies get within a certain user-defined distance.
\end{itemize}}

\subsection{State Estimation}

{In addition to the numerical propagation modules of Tudat, state estimation (using a batch least square filter) can also be performed. This functionality was used by \cite{BauerEtAl2014} for the orbit determination for LRO using laser ranging data. Simulation models for various astrometric and radiometric data types are available, and applied by \emph{e.g.} \cite{DirkxEtAl2014, DirkxEtAl2016b, DirkxEtAl2017, DirkxEtAl2018}, with recent publications focussing on the JUICE-PRIDE radio tracking experiment \citep{GurvitsEtAl2013}. Models for the detailed analysis of real radio and optical data are currently under development.}

{In addition to the settings listed above for the numerical propagation, Tudat requires the following information to perform a state estimation:
\begin{itemize}
\item A list of parameters to estimate, which may include initial rotational and/or translational state (single-arc, multi-arc, or a combination of the two), gravity field coefficients, Parametric Post-Newtonian (PPN) parameters, \emph{etc.} A user may, but need not, provide an \emph{a priori} covariance matrix for these parameters.
\item Observation models and settings. Various types of observation models (such as range, range-rate, angular position) can be modelled by Tudat, where a user has the freedom to add bias models, atmospheric, relativistic corrections, and other settings that may be needed, such as integration time for closed-loop Doppler observables.
\item Observations and associated weights. The input for the estimation is a list of realizations of observations, with associated time tag and weights. These observations may be simulated by Tudat, or obtained from data archived (\emph{e.g.} PDS) or other simulation tools.
\end{itemize}
Using the above information, acceleration and observation partial derivative models are automatically set up, and the associated variational equations for $\Phi(t,t_{0})$ and $S(t)$ are solved numerically.}

\subsection{External Libraries}

{In addition to its own codebase, Tudat links to a number of external software libraries, primarily:
\begin{itemize}
\item Boost: The Boost libraries\footnote{\url{https://www.boost.org/}} are a collection of C++ libraries, which is used in support of various Tudat functions, primarilly unit testing, multi-dimensional arrays, file reading/writing and file system acces.
\item CSPICE: The SPICE toolbox\footnote{\url{https://naif.jpl.nasa.gov/}} \citep{Acton1996} is used in Tudat primarilly to retrieve ephemerides, rotational states, and other physical quantities, such as gravitational parameters and radii of solar system bodies.
\item SOFA: The Standards of Fundamental Astronomy (SOFA) library\footnote{\url{http://www.iausofa.org/}} is used in Tudat primarily for Earth orientation and time conversion functionality
\item Eigen: Tudat uses the Eigen library\footnote{\url{https://eigen.tuxfamily.org/}} for all its linear algebra operations
\item Pagmo2:  Tudat provides an optional interface to the Pagmo2 software library\footnote{\url{https://esa.github.io/pagmo2/}}, a global optimization toolbox developed by ESA's Advanced Concept Team (ACT).
\end{itemize}
When cloning our "Tudat bundle" repository\footnote{\texttt{\url{https://github.com/tudat/tudatBundle}}}, Tudat and all other required libraries are automatically downloaded.}

\end{appendix}

\bibliographystyle{apalike} 
\bibliography{Bibliography}

\end{document}